\documentclass[aps,superscriptaddress,twocolumn]{revtex4-1}
\usepackage[utf8]{inputenc}

\usepackage{graphicx}
\graphicspath{{Figures/}}

\usepackage{array}

\usepackage{amsmath}

\usepackage{tabularx}

\begin{document}

\title{On topological defects in two-dimensional orientation-field models for grain growth}

\newcommand\WIGNER{Institute for Solid State Physics and Optics, Wigner Research Centre for Physics, PO Box 49, H-1525 Budapest, Hungary}
\newcommand\EP{Physique de la Matière Condensée, École Polytechnique, CNRS 91128, Palaiseau, France}
\newcommand\NIST{National Institute of Standards and Technology, Gaithersburg, MD 20899, US}

\author{Bálint Korbuly}
\affiliation{\WIGNER}
\author{Mathis Plapp}
\affiliation{\EP}
\author{Hervé Henry}
\affiliation{\EP}
\author{James A. Warren}
\affiliation{\NIST}
\author{László Gránásy}
\affiliation{\WIGNER}
\author{Tamás Pusztai}
\affiliation{\WIGNER}

\date{\today}

\begin{abstract}
Standard two-dimensional orientation-field based phase-field models rely on a continuous scalar field to represent crystallographic orientation. The corresponding order parameter space is the unit circle, which is not simply-connected. This topological property has important consequences for the resulting multi-grain structures: (i) trijunctions may be singular; (ii) for each pair of grains, there exist two different grain boundary solutions that cannot continuously transform to one another; (iii) if both solutions appear along a grain boundary, a topologically stable, singular point defect must exist between them. While (i) can, (ii) and therefore (iii) cannot be interpreted in the classical picture of grain boundaries. In addition, singularities cause difficulties, such as lattice pinning in numerical simulations. To overcome these problems, we propose two new formulations of the model. The first is based on a 3-component unit vector field, while in the second we utilise a 2-component vector field with an additional potential. In both cases, the additional degree of freedom introduced make the order parameter space simply-connected, which removes the topological stability of these defects.
\end{abstract}


\maketitle

\section{Introduction}
\label{sec:introduction}

Polycrystalline materials are solids that consist of small, differently oriented crystallites, called grains. The regions where the grains meet are the grain boundaries. In the simplest picture they are considered as thin transient zones between the neighbouring grains which have perfect crystalline order and well defined crystallographic orientation. Polycrystalline materials are usually formed by the freezing of their melt, a process in which new crystallites nucleate, grow and impinge on each other. After full solidification the growth of the grains can continue on the expense of each other. This grain coarsening process is governed by the minimization of the grain boundary area, as it decreases the excess energy due to the grain boundary network.

The phase-field method is a very powerful tool for modeling solidification~\cite{Boettinger2002, Hoyt2003, Pusztai2008, Asta2009, Steinbach2009, Provatas2010, Steinbach2013, Granasy2014}, including the nucleation and the subsequent growth of a solid phase in its melt. There are two very distinct approaches to address polycrystals in the phase field theory. The first approach is to use the multi-order-parameter~\cite{Chen1994, Moelans2008} or multi-phase-field~\cite{Steinbach2009, DarvishiKamachali2012, Toth2015} models that assign separate order parameters to different grains. These order parameters or phase-fields are constant inside the grains and change continuously through the grain boundary. The other approach is to keep a single order parameter, the phase-field to represent the crystallinity of the material, and to add a new field, the orientation-field to represent the local crystalline orientation~\cite{Warren1998, Kobayashi2000, Warren2000, Granasy2002, Warren2003, Granasy2003, Granasy2004, Granasy2005, Henry2012}. In two dimensions (2D) there is one orientational degree of freedom, which is usually represented by a single scalar field, while in three dimensions (3D) there are three orientational degrees of freedom, and more complex constructions, such as quaternions or rotation matrices are used to represent them~\cite{Pusztai2005, Kobayashi2005}.

Both approaches have their advantages and disadvantages. The multi-phase-field models require a large number order parameters, usually considered as $N$-component vectors that represent either $N$ grains (all grains can have different orientations) or $N$ distinct orientations (all grains can have one of these $N$ orientations). Even in the latter case, a large number of fields is required. Fortunately, optimization techniques exist that reduce the number of fields one really has to compute in a region of a simulation to a few~\cite{Vedantam2006, Gruber2006, Vanherpe2007}. In contrast, the orientation-field models with their one or very few extra fields seem to be inherently more efficient.

In the present work, we focus on the 2D orientation field models. Using the terminology of Ref.~\cite{Mermin1979}, we consider the polycrystalline structure as an ordered medium which is described by an order parameter field $\theta(\mathbf{r})$ that assigns an orientation to every point of the 2D space. A general 2D orientation may take non-equivalent values from an interval of length $2\pi$, e.g.\ $\theta \in [0, 2\pi[$ with the end points being equivalent. Crystal structures may have additional n-fold rotational symmetries, which decrease this interval to $\theta \in [0, 2\pi/n[$. However, as by a simple rescaling of $\theta$, which does not effect the topological properties, the n-fold symmetric case can be mapped to the general one, we can assume that our system has no extra rotational symmetries. This scenario is equivalent to the case of planar spins discussed in Ref.~\cite{Mermin1979}. Such a field can exhibit \emph{topological defects}: consider a closed loop in space, and follow the orientation along the loop. If its total increment is non-zero, there is a topological defect within the loop. Well-known examples are the ``hedgehog'' pattern of electrical field lines surrounding an isolated charge, or the ``triangles'' and ``U-turns'' that you can find in the line patterns on your fingertips.

\begin{figure}
\centerline{\includegraphics[width=\linewidth]{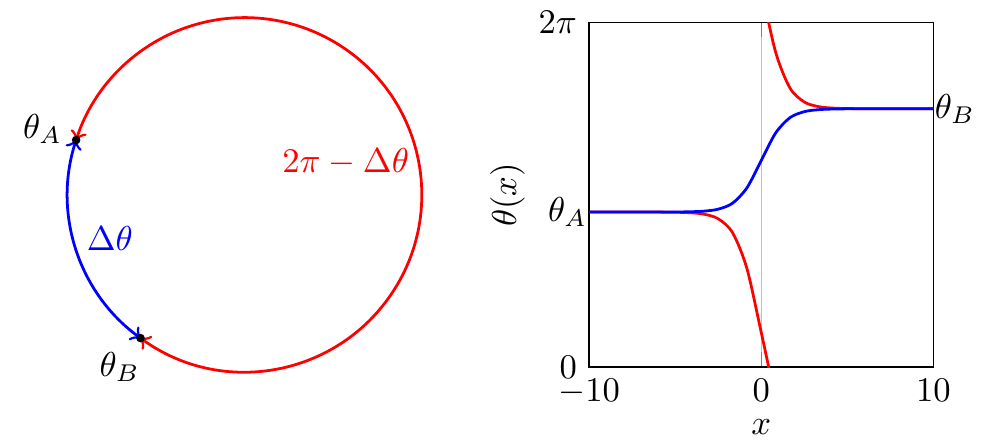}}
\begin{tabularx}{\linewidth}{XX}
\centering (a) & \centering (b)
\end{tabularx}
\caption{Due to the $2\pi$ periodicity of the orientation field $\theta$, there exist two continuous paths connecting the orientations $\theta_A=0.9\pi$ and $\theta_B=1.6\pi$ (a). The ``small turn'' connection is shown by the blue, while the ``large turn'' connection is shown by the red line. These paths correspond to two different grain boundary solutions (b). Due to topological reasons, there is no continuous transformation between these two paths/solutions. Please note that as $0$ and $2\pi$ are equivalent orientations, the red profile on the right is also continuous.}
\label{fig:twosolutions}
\end{figure}

In orientation-field models, the energy penalty for orientation variations inside the bulk solid is high, and therefore such configurations are never observed. However, topological defects may be ``hidden'' in grain boundaries or trijunctions where the strong variations of the orientation are localized. We will mainly focus on topological defects in grain boundaries. The order parameter space of the model, that is, the set of all possible values of the order parameter, is the unit circle, where the term circle is used in its strict technical sense, meaning the 1-sphere or in more common words, the circumference of the unit disk. Consider two grains of orientations $\theta_A$ and $\theta_B$. As depicted in Fig.~\ref{fig:twosolutions}, there are two ways to connect these two orientations. These two paths are topologically distinct, because they cannot continuously be deformed one into the other. If two parts of the same grain boundary are occupied by the two different solutions, a topological defect is present (for a more detailed description, see further below). We have found in numerical simulations that such defects can indeed form during the ``natural'' evolution of the grain boundary network.

This raises two problems. First, no microscopic interpretation (in terms of atomistic configurations) can be given to such defects, which makes them seem to be artifacts of the continuum orientation field formulation. Second, since the orientation field is singular around such defects, lattice pinning of grain boundaries may occur during numerical simulations, which alters the grain boundary dynamics. It is actually through this effect that we have first noticed the presence of topological defects in grain boundaries. Consequently, they are an undesirable feature of the model and should be eliminated.

We present two different ways to achieve this goal, which are based on topological arguments. The stability of the defects can be linked to the fact that the order parameter space of the model, the unit circle, is not simply connected. Consequently, we extend the model by replacing the scalar orientation field either by a three-component unit vector or by a two-component vector without length constraint, both of which have simply-connected order parameter spaces. We demonstrate that, in both models, the ``longer'' path in Fig.~\ref{fig:twosolutions} becomes unstable and is eliminated. Consequently, the topological defects also disappear.

In the following, we will first recall some fundamentals of topology, and then expound the consequences of using the standard 2D orientation field in Section~\ref{sec:problemdetail}. Some of them are identified as problems when numerical simulations of grain growth are considered. Section~\ref{sec:models} is dedicated to the description of the original and the proposed two new formulations of the orientation-field-based phase-field models. In Section~\ref{sec:results}, we show how the problems identified appear in the original model and how they are cured in the new models. We close the paper by a summary in Section~\ref{sec:summary}.

\section{Detailed description of the problem}
\label{sec:problemdetail}

In this section, we make a detailed exploration of the consequences of using a continuous scalar field as an orientational order parameter in 2D. We address three phenomena: the singularity of trijunctions, the existence of two different grain boundary solutions for the same pair of grains and the appearance of topological point defects on the grain boundaries. 

\subsection{Background}

Reference~\cite{Mermin1979} reviews general properties of spatially extended systems that are described by order parameter fields of various nature. For models with a scalar orientation field, in which two angles separated by a multiple of $2\pi$ are the same, the order parameter space can be visualized as the unit circle. This is a one-dimensional (1D) space which is not simply-connected. This means, by definition, that there must exist a loop in it which cannot continuously shrink to a point~\cite{Mermin1979}. In our case this loop is the circle itself.

\begin{figure}
\includegraphics[width=0.5\linewidth]{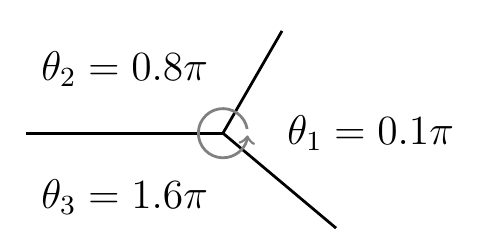}\includegraphics[width=0.5\linewidth]{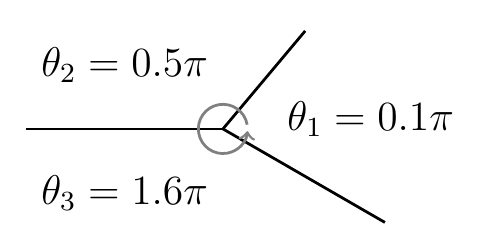}
\begin{tabularx}{\linewidth}{XX}
\centering (a) & \centering (b)
\end{tabularx}
\caption{Schematic view of trijunctions where the orientation field is singular (a) / non-singular (b). The solid lines may equally correspond to abrupt changes in the orientation (such as in sharp interface models) or may indicate thin regions where the change in orientation is continuous (such as in orientation-field based phase-field models). The sum of orientational increments around the trijunctions are $(0.7+0.8+0.5)\pi = 2\pi$ (winding number = 1) and $(0.4-0.9+0.5)\pi = 0$ (winding number = 0), respectively. As the winding number is an integer quantity, it cannot be changed by continuous transformations in the orientation field.}
\label{fig:trijunction}
\end{figure}

In general, even using a model which produces continuous fields, there may be isolated regions of the physical space where the order parameter field is non-continuous. These singular regions are called defects. It is of fundamental importance to distinguish defects that can be eliminated by ``local surgery''~\cite{Mermin1979}, i.e., by continuous changes of the order parameter field in the neighborhood of the defect, from those that cannot. Defects belonging to the first type are called topologically unstable and can be eliminated by continuous models. Our relevant example is the simple grain boundary. In the sharp interface description the orientation changes abruptly between grains, but if the orientation-field is made continuous just by smoothing it out in a narrow region around the grain boundary, this singularity is removed. This is precisely what orientation-field-based phase-field models do~\footnote{The only exceptions are the special version of the Kobayashi-Warren-Carter model and its descendants, which use a single $|\nabla\theta|$ term in the free energy functional. In theory, this version produces non-continuos step-like orientation field at grain boundaries. In general numerical simulations, however, pixels with intermediate values appear.}, therefore these line singularities do not appear in them. In contrast, point defects belonging to the second type may appear even in (or exclusively in) models that use a continuous orientation field. Depending on the relative orientation of the neighboring grains, trijunctions may serve as a simple examples for singular points (see Fig.~\ref{fig:trijunction} and the detailed description later in Section~\ref{sec:problemdetail}). The point defects may be classified by their winding number, which is defined as the number of revolutions the order parameter makes as one travels along a path encircling the defect once in the positive, counterclockwise direction. We would like to recite two important statements of topology~\cite{Mermin1979} that helps us understanding the behavior of defects. The first is that defects with the same winding number can, while defects with differing winding numbers cannot be continuously transformed into each other. As a consequence, an isolated defect with nonzero winding number is topologically stable. The second statement is that a pair of defects is topologically equivalent to a single defect with winding number equal to the sum of the winding numbers of the individual defects. This means, e.g., that two defects with winding numbers $+1$ and $-1$ can annihilate.

\subsection{Singular trijunctions}
\label{sec:problem0}

In 2D, trijunctions are points where three neighboring grains meet (Fig.~\ref{fig:trijunction}). We do not explicitly address quad- or even higher multi-junctions as they do not normally appear in 2D polycrystalline structures. If we take a circular path once around the trijunction in the positive direction and add up the increments of the (scalar) orientation field along this path, we must end up with an integer number of revolutions, called the winding number. As this number could change only by discrete steps, it has to remain constant as we continuously decrease the radius $r$ of this circular path, supposed that the orientation field is continuous at least outside the trijunction point. For nonzero winding numbers this also means that the orientation field is singular because its directional derivatives along the circular path must diverge in the $r \rightarrow 0$ limit. As shown by its nonzero winding number, this singularity is topologically stable.

\subsection{Two different grain boundary solutions}
\label{sec:problem1}

Let us consider two neighboring grains with orientations $\theta_A$ and $\theta_B$. These orientations correspond to two points in the order parameter space. Any continuous path between the two grains in the real space maps to a continuous path between the respective two points in the order parameter space. In our case, when the order parameter space is a circle, $\theta_A$ and $\theta_B$ can be connected by two different paths, see Fig.~\ref{fig:twosolutions}. One is usually shorter, corresponding to a smaller turn by $\Delta\theta$ in one, the other is usually longer, corresponding to a larger turn by $2\pi-\Delta\theta$ in the other direction. These two solutions correspond to two different grain boundaries which, in general, have different energies. It is important to stress that the two solutions cannot be transformed to one another with continuous transformations, e.g.~by models such as the phase-field models we consider.

\subsection{Topological defects at grain boundaries}
\label{sec:problem2}

\begin{figure}
\includegraphics[width=\linewidth]{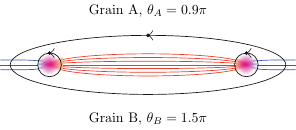}
\caption{Schematic view of a grain boundary containing two defects where the ``small turn'' and ``large turn'' solutions meet. The two different grain boundary profiles are represented by the blue and red isolines that correspond to steps of $0.2\pi$ in $\theta$. In the magenta region the orientation field is significantly distorted as the different grain boundary profiles try to match each other, but still assumed to be continuous. The integral of $\nabla\theta$ along the black paths are $+2\pi$ ($-2\pi$) for the left (right) circle, and $0$ for the ellipse. This means that the magenta regions must contain topologically stable defects with winding number $+1$ and $-1$, but the larger area which includes both defects can be made defect free.}
\label{fig:defectsatgb}
\end{figure}

Let us assume that both solutions discussed in Section~\ref{sec:problem1} appear along the same grain boundary. Figure~\ref{fig:defectsatgb} shows a relaxed {\it XYX} sandwich structure, where {\it X} and {\it Y} stands for vertical slabs of the matter which contain a horizontal grain boundary with orientation profiles corresponding to the ``small turn'' ({\it X}, shown in blue) and ``large turn'' ({\it Y}, shown in red) solutions, respectively. As described in the figure caption, each magenta region where the different types of solutions merge must contain a defect with respective winding number $+1$ and $-1$. As will be shown in Section~\ref{sec:results} (see e.g.~Figure~\ref{fig:pinning1c2}) similar structures do appear in real simulations. As their nonzero winding numbers indicate, these defects are topologically stable, meaning that they cannot disappear by continuous changes of the orientation field inside the black circles. The two defects become topologically unstable, however, if we consider them together in a larger area that includes both of them, e.g., inside the black ellipse, as the respective winding number is zero.

Topological instability is a necessary, but not sufficient condition for these defects to disappear. If it is allowed topologically, other factors, such as energetics come into picture. Let us consider a system of two defects similar to the setup shown in Figure~\ref{fig:defectsatgb}. If $\theta_A$ and $\theta_B$ were opposite points of the order parameter space, then the two solutions would be symmetric and the two types of grain boundaries would have the same energy. In this case there would be no driving force for the defects to move, they would stay in their neutral equilibrium positions. However, in the general asymmetric case the two types of grain boundaries have different energies and there is a driving force for the defect to move in the direction that makes the lower energy grain boundary longer on the expense of the higher energy grain boundary. In spite of their singularity, the movement of the defects can happen via continuous changes of the orientation field, they can drift as the driving force requires. The defects shown in Fig.~\ref{fig:defectsatgb} would drift towards each other and annihilate, resulting in a defect-free final setup. In contrast, if we started from a {\it YXY} instead of the {\it XYX} sandwich structure, the defects would drift away from each other and therefore they could not annihilate.

\bigskip

After this overview of the potential issues, we should discuss their relevance to grain growth in real materials. It is important to stress that all of them are the direct consequence of using a continuous 2D scalar orientation field for describing a polycrystalline structure. This means that if we have a model that uses a scalar order parameter for the orientation, we can expect these phenomena to appear. It also means that if we consider the continuous scalar orientation field as a good description of multi-grain structures, then we have to accept its consequences as well. 

First, let us consider the simplest classical, continuum picture of a grain boundary, where it is considered as a thin, continuous transient zone between two homogeneous grains. The excess energy of this transient zone, i.e., the grain boundary energy depends on the misorientation of the grains and the inclination of this interface. If, for simplicity, we take the isotropic approximation by ignoring the inclination dependence, then the grain boundary energy becomes the function of the misorientation only. For a small angle grain boundary we might imagine a small, continuous turn of some locally defined orientation by $\Delta\theta$, but its pair, the nearly full but still continuous turn by $2\pi-\Delta\theta$ is not a good picture of the local structure. Also, allowing for the two different solutions would mean that a grain boundary can have two different energies for the same misorientation, and would immediately allow for the appearance of topological defects on the grain boundaries with their associated singularity. These are not included in the simplest classical picture. By physical sense, the solution for all these problems is to omit the higher energy profiles, or ideally, construct a model in which they do not appear. To our knowledge, the only work in this direction is Ref.~\cite{Warren2003}, where to eliminate the higher energy profiles a special correction procedure was executed after every 100th--1000th steps of their simulations. In our approach, this extra procedure is not required, as the models are constructed in a way that the high energy profiles lose their topological stability and can transform to the low energy ones during the normal course of time evolution.

Second, on the atomistic scale, defects can be identified at grain boundaries. For 2D systems, defects have been discussed e.g.\ in polycrystalline graphene~\cite{Yazyev2014}, the Ginzburg--Landau type model for diblock copolymers~\cite{Vega2005}, the Phase-Field Crystal model~\cite{Granasy2016} and molecular dynamics simulations~\cite{Sun2016}. Since the ordered state in these systems is hexagonal, grain boundaries consist of dislocations and disclinations. Disclinations are defects at atoms that have 5 or 7 nearest neighbors, as opposed to 6 corresponding to the regular triangular lattice. Dislocations can be considered as pairs of disclinations/atoms with 5 and 7 nearest neighbors. More generally speaking, according to the continuum theory of defects (see e.g.~ \cite{Kroener}) grain boundaries in polycrystals (of any crystallographic structure) can always be described as arrays of dislocations. It might therefore be tempting to think that the defects appearing in the orientation field models could correspond to some real defects. However, in a continuum theory dislocations and disclinations are singularities of the displacement field. Those can only be stable in presence of a discrete translational symmetry due to the existence of objects or domains of a characteristic scale. Their topological charge is a vector (the Frank vector for disclinations and the Burgers vector for dislocations). In contrast, in orientation-field models there is only an orientational order, and the topological charge is a scalar. Therefore, the defects of the orientation-field models cannot correctly match with the description of dislocations or disclinations.

As a bottom line, we stay with the simple classical picture in this paper and consider the resulting defects as problems that we should eliminate. Our goal is therefore to ``patch'' the orientation field model to be free of the above problems. The key step in our approach is to choose new order parameters to represent the 2D orientation which has a simply-connected order parameter spaces. This way the topological stability of the defects can be removed, allowing them to disappear completely where needed (defects on grain boundaries) or just become nonsingular (trijunctions).

\section{The models}
\label{sec:models}

We restrict our work to the 2D orientation-field based phase-field models, where the structural order parameter, the $\phi(\mathbf{r})$ phase-field is coupled to some $\theta(\mathbf{r})$ orientation-field that represents orientation in 2D. In almost all previous works $\theta(\mathbf{r})$ is a scalar field which gives the 2D orientation as an angle in a reference frame. Two main formulations of this approach exist in the literature. Their distinctive feature is how they attain localized grain boundaries. In the earlier Kobayashi-Warren-Carter (KWC) formulation~\cite{Kobayashi2000}, a term proportional to $|\nabla\theta|$ is added to the usual $|\nabla\theta|^2$ term in the free energy density, while in the later Henry-Mellenthin-Plapp (HMP) model~\cite{Henry2012}, only the $|\nabla\theta|^2$ term is used but with a singular coupling function $g(\phi)$. In spite of this important difference, the two models can produce very similar results both for polycrystalline solidification and grain coarsening~\cite{Korbuly2017a,Korbuly2017b}. As they both rely on a continuous scalar orientation field, they both suffer from the problems introduced earlier. Since from our viewpoint there is no real difference between them, we chose the HMP model for our study. Furthermore, as our goal was just to illustrate the topological problems and to show how to overcome them, we did not attempt to model any specific material and therefore we used the non-dimensional forms of the model equations with parameters in the order of unity.

For reference, we recite the main equations of the HMP model from Ref.~\cite{Henry2012} in Subsection~\ref{sec:oldmodel}. In Subsection~\ref{sec:equivmodel} we present a mathematically equivalent formulation of this original model, where a two-component unit vector field is used instead of the scalar $\theta(\mathbf{r})$. We do this, because the two new models we propose in Subsections~\ref{sec:3compmodel} and~\ref{sec:2compmodel} are more easily introduced as the extension of this equivalent formulation than based on the original HMP model.

\subsection{The original model}
\label{sec:oldmodel}

In the original HMP model~\cite{Henry2012} the total free energy of the system is a functional of the $\phi$ and $\theta$ fields,
\begin{align}
\begin{split}
\label{eq:originalf}
F[\phi,\theta] &= \int f(\phi,\nabla\phi,\nabla\theta) \, dV = \\
&= \int \left[ \frac{1}{2} (\nabla\phi)^2  + V(\phi) + f_\mathrm{ori}(\nabla\theta) \right] dV,
\end{split}
\end{align}
where $f$ is the local free energy density. The 
\begin{equation}
V(\phi) = \phi^2(1-\phi)^2 - u \lambda \, \phi^3 (6\phi^2-15\phi+10) \\
\end{equation}
potential includes the usual double-well and tilt functions, with the combination $u\lambda$ being the non-dimensional driving force for solidification. A small difference compared to the original HMP model is that we used the more traditional form of the tilt function instead of the one in Ref.~\cite{Henry2012}. The term which is of most interest for us is the contribution of the orientation field to the free energy density, 
\begin{equation}
f_\mathrm{ori}(\nabla\theta) = \mu^2 g(\phi) (\nabla\theta)^2,
\label{eq:foriHMP}
\end{equation}
where $\mu^2$ is the strength of the coupling and 
\begin{equation}
g(\phi) = \frac{7\phi^3-6\phi^4}{(1-\phi)^3}
\end{equation}
is the singular coupling function (corresponding to $\alpha=3$ in Ref.~\cite{Henry2012}) specific to the HMP model. 

In equilibrium the variational derivative of $F$ with respect to the fields $X=\phi,\theta$ has to be zero,
\begin{equation}
\label{eq:elgeneral}
\frac{\delta F}{\delta X} = \frac{\partial f}{\partial X} - \nabla \frac{\partial f}{\partial \nabla X} = 0,
\end{equation}
which define the 
\begin{align}
\label{eq:elphi}
V'(\phi) + \mu^2 g'(\phi) (\nabla\theta)^2 - \nabla^2\phi = 0 \\
\label{eq:eltheta}
\mu^2 \nabla( g(\phi) \nabla\theta) = 0
\end{align}
Euler-Lagrange equations. Out of equilibrium, the time evolution of the system is assumed to follow the standard variational dynamics, which, for these non-conserved order parameters, result in the 
\begin{align}
\label{eq:phidot0}
\dot{\phi} &= -M_\phi \frac{\delta F}{\delta \phi} = M_\phi \left( \nabla^2\phi - V'(\phi) - \mu^2 g'(\phi) (\nabla\theta)^2 \right)\\
\label{eq:thetadot0}
\dot{\theta} &= -M_\theta(\phi) \frac{\delta F}{\delta \theta} =  M_\theta(\phi) \mu^2 \nabla( g(\phi) \nabla\theta)
\end{align}
equations of motion. Here, $\dot{\phi}$ and $\dot{\theta}$ stand for the time derivatives, $M_\phi$ and $M_\theta(\phi)=1/g(\phi)$ are the mobilities of the phase field and the orientation field. The particular form of $M_\theta(\phi)$ was chosen to counterbalance the divergence of $g(\phi)$ that multiplies $\nabla\theta$ in Eq.~\ref{eq:thetadot0}.

Unfortunately, analytic solutions of these partial differential equations are limited to the simplest case of an equilibrium grain boundary in 1D. Even then, the profiles corresponding to a general misorientation $\Delta\theta$ cannot be provided in closed form. Therefore in all practical cases we need to rely on numerical solutions. All numerical results presented in this paper were obtained by solving the equations of motion of the respective models (Eq.~\ref{eq:phidot0} and~\ref{eq:thetadot0} for the original HMP model) by a simple finite differencing and forward Euler stepping scheme. This includes the equilibrium solutions, which we determined as the long-time stationary solutions of the dynamical equations (instead of solving the respective Euler-Lagrange equations directly).

Please note, that this simple formulation is isotropic in the sense that the energy of an interface does not depend on its inclination. The only parameter the grain boundary energy depends on is the misorientation of the grains. This simplification does not alter the general topological considerations and the conclusions of our work.

A final important comment regarding the numerical simulations: since $\theta$ is an angular representation of the 2D orientation, the metric used by the gradient operator in $\nabla\theta$ is the difference of angles. The difference of $\theta_B$ and $\theta_A$ is the angle of the rotation which transforms the orientation $\theta_A$ to $\theta_B$, and it is equivalent to the directed distance of the respective points in the order parameter space, i.e., along the unit circle (see Fig.~\ref{fig:twosolutions}). When calculating this difference the $2\pi$ periodicity of $\theta$ has to be taken into account. It means that from the two possible rotations the one with smaller magnitude has to be chosen, which corresponds to the shorter path along the unit circle. For a possible implementation of this procedure see Ref.~\cite{Warren2003}.

\subsection{An equivalent formulation of the original model}
\label{sec:equivmodel}

An equivalent formulation of the original model can be obtained by using a 2-component unit vector ${\boldsymbol{\theta}}=(\theta_1,\theta_2)$ instead of its polar angle $\theta$ (as in the HMP model) to represent the 2D orientation. This representation has the same order parameter space, the unit circle. The orientational part of the free energy density is changed to
\begin{equation}
\label{eq:fori1}
f_\mathrm{ori}(\nabla\boldsymbol{\theta}) = \mu^2 g(\phi) \sum_{i=1}^2 (\nabla\theta_i)^2,
\end{equation}
but everything else remained unaltered.

Now we have to deal with two scalar fields instead of one, but with a constraint $\theta_1^2+\theta_2^2=1$ between them. This constraint is taken into account by the standard Lagrange multiplier method when deriving the equations of motion for this model. The detailed calculation for the general $N$-component case is shown in the Appendix, here we just show the results for $N = 2$ and the free energy functional given by Eq.~\ref{eq:originalf} and~\ref{eq:fori1}:

\begin{align}
\label{eq:phidot1}
\dot{\phi} &= M_\phi \left( \nabla^2\phi - V'(\phi) - \mu^2 g'(\phi) \sum_{i=1}^2 (\nabla\theta_i)^2 \right)\\
\label{eq:thetadot1}
\dot{\theta_i} &=  M_\theta(\phi) \mu^2 \left( \nabla( g(\phi) \nabla\theta_i) - \theta_i \sum_{k=1}^2 \theta_k \nabla( g(\phi) \nabla\theta_k) \right).
\end{align}

Though this model is mathematically equivalent to the original HMP model, there is a slight difference between them in numerical simulations. This is related to the change of metric in the order parameter space. As noted in the previous subsection, the original HMP model relies on the difference of \emph{angles}, which corresponds to the \emph{arc length}, while this unit vector model relies on the usual \emph{euclidean distance} which corresponds to the \emph{chord} between the respective points of the unit circle. For small differences in the orientations that we expect in a well-resolved numerical simulation one is a good approximation of the other, making the two models nearly identical. For infinitesimal differences the arc is the same as the chord and therefore the two models are equivalent.

\subsection{The unit sphere model}
\label{sec:3compmodel}

As discussed in the Introduction, the problem of the above models originate in the fact that their order parameter space is not simply-connected. This suggests that to overcome these problems we should choose a new representation which has a simply-connected order parameter space. A straightforward approach is to extend the order parameter space in the third dimension, allowing the order parameter to take values that correspond to points on the surface of an unit sphere instead of to points on the unit circle. Thus the name unit sphere (US) model. This opens the possibility of transforming the two different continuous connections between $\theta_A$ and $\theta_B$ (red and blue lines in Figure~\ref{fig:twosolutions}) continuously into one another, just as an elastic band with fixed ends at $\theta_A$ and $\theta_B$ could be moved between the red and the blue arcs, if sliding on the surface of a sphere is allowed.

To this change of the order parameter space there corresponds the generalization of the orientation field to a 3-component unit vector, ${\boldsymbol{\theta}}=(\theta_1,\theta_2,\theta_3)$. Just as in the previous 2-component model, the ``true'' scalar orientation $\theta$ that finally represents the crystallographic orientation is the polar angle defined by the components $\theta_1$ and $\theta_2$. The third component is best considered as an additional degree of freedom that allows to overcome the topological limitations of the original model, where necessary. The corresponding free energy functional and equations of motion are also the straightforward generalization of the 2-component unit vector model to $N=3$,
\begin{equation}
\label{eq:fori2}
f_\mathrm{ori}(\nabla\boldsymbol{\theta}) = \mu^2 g(\phi) \sum_{i=1}^3 (\nabla\theta_i)^2
\end{equation}
and 
\begin{align}
\label{eq:phidot2}
\dot{\phi} &= M_\phi \left( \nabla^2\phi - V'(\phi) - \mu^2 g'(\phi) \sum_{i=1}^3 (\nabla\theta_i)^2 \right)\\
\label{eq:thetadot2}
\dot{\theta_i} &=  M_\theta(\phi) \mu^2 \left( \nabla( g(\phi) \nabla\theta_i) - \theta_i \sum_{k=1}^3 \theta_k \nabla( g(\phi) \nabla\theta_k) \right).
\end{align}

A small, but important detail must be emphasized, though. If $\theta_3=0$ everywhere (a natural choice for the initial conditions) then $\dot\theta_3=0$, too, meaning that $\theta_3$ will remain zero. In this limit the model is exactly the same as the previous 2-component unit vector model, which is equivalent to the original HMP model. Therefore all these models share the same grain boundary solutions and grain boundary properties. However, the unphysical, but topologically stable solutions of the HMP model are expected to be unstable solutions of this model. These solutions, by applying a small perturbation to $\theta_3$ will  transform to other, stable solutions. We can make it very intuitive using our simple mechanical analogy. A rubber band with fixed ends along the red arc of the equator as shown in Fig.~\ref{fig:twosolutions} is in unstable equilibrium. If a small perturbation is applied, it flips to the opposite blue arc, which corresponds to its stable equilibrium position on the sphere.

\subsection{The Landau--De Gennes model}
\label{sec:2compmodel}

In this model, the simple-connectedness of the order parameter space is achieved by extending the circular order parameter space to the whole plane embedding the circle. This is attained by using the same 2-component vector orientation field as in Section~\ref{sec:equivmodel}, but replacing the hard constraint $\theta_1^2+\theta_2^2=1$ with a soft constraint that allows all points of the plane, but still prefers the unit circle. For this, we added a new term, a sombrero-shaped potential
\begin{equation}
\label{eq:sombrerofun}
f_s({\boldsymbol\theta}) =  \left(1-\boldsymbol{\theta}^2\right)^2 = \left( 1-\sum_{i=1}^2 \theta_i^2 \right)^2,
\end{equation}
to the free energy density, which has global minima at $|{\boldsymbol{\theta}}|=1$ and local maximum at $|{\boldsymbol{\theta}}|=0$ and is very similar to the Landau--De Gennes potential used for the description of nematic liquid crystals~\cite{Gennes1995}. Due to this similarity, we call this model the Landau--De Gennes (LDG) model.

The orientation part of the free energy density is therefore
\begin{equation}
\label{eq:fori3}
f_\mathrm{ori}(\nabla\boldsymbol{\theta},\boldsymbol\theta) = \mu^2 g(\phi) \left[ \sum_{i=1}^2 (\nabla\theta_i)^2 + \nu f_s({\boldsymbol\theta}) \right],
\end{equation}
where $\nu$ sets the strength of the new potential. Large $\nu$ values are expected to keep $|{\boldsymbol{\theta}}|$ close to $1$, approximating the 2-component unit vector model, while small values of $\nu$ make the system softer, allowing $|\boldsymbol{\theta}|$ deviate from $1$ significantly, and also making the transition through the barrier centered at the origin easier. 

Deriving the equation of motion for $\phi$ is straightforward. For $\boldsymbol{\theta}$, we use the non-constrained equation of motion (Eq.~\ref{eq:standardeom} in the Appendix), but including the new potential. Finally, we obtain
\begin{align}
\label{eq:phidot3}
\begin{split}
\dot{\phi} &= M_\phi \Bigg( \nabla^2\phi - V'(\phi) - \\
 &- \mu^2 g'(\phi) \left[ \sum_{i=1}^2 (\nabla\theta_i)^2 + \nu \left( 1-\sum_{i=1}^2 \theta_i^2 \right)^2 \right] \Bigg) \\
\end{split} \\
\begin{split}
\label{eq:thetadot3}
\dot{\theta_i} &=  M_\theta(\phi) \mu^2 \left[ \nabla( g(\phi) \nabla\theta_i) - 4 \nu g(\phi) \theta_i \left( 1-\sum_{i=1}^2 \theta_i^2 \right) \right]
\end{split}
\end{align}
as equations of motion for this model.

\section{Results and Discussion}
\label{sec:results}

In this section we present numerical simulations that illustrate how the new models proposed in Section~\ref{sec:3compmodel} and~\ref{sec:2compmodel} overcome the problems of traditional models with scalar valued orientation field. For reference, we first show the results obtained by the original HMP model. In all examples shown we used the following dimensionless parameters: $u\lambda = 0$, $\mu = 1/(2\pi)$, $M_\phi = M_\theta = 1$, $\Delta x = 0.05$, $\Delta t = 0.00025$.

\subsection{Elimination of the unphysical grain boundary solutions}
\label{sec:UGBelimination}

First, 1D simulations were carried out to determine the grain boundary solutions of the original and the two newly proposed models. The orientations of the neighboring grains were set to $\theta_A=0.6\pi$ and $\theta_B=1.4\pi$ in the examples below. The two possible continuous transitions from grain~$A$ to grain~$B$ correspond to the orientation field is either gradually \emph{increasing} by $\Delta\theta_1=0.8\pi$ or gradually \emph{decreasing} by $\Delta\theta_2=1.2\pi$ through the grain boundary. The deviation from the symmetric configuration can be measured by the parameter $\epsilon = |\Delta\theta_2-\Delta\theta_1|/2$.

\subsubsection{Results of the HMP model}

The reference model was solved by simulating Eq.~\ref{eq:phidot0} and~\ref{eq:thetadot0} on the interval from $x_A=-5$ to $x_B=5$ with boundary conditions $\phi^\prime(x_A)=0$, $\theta(x_A)=\theta_A$ on the left and $\phi^\prime(x_B)=0$, $\theta(x_B)=\theta_B$ on the right ends. To obtain the two different solutions, two different initial conditions were used. For $\theta(x)$ we chose profiles that changed between the end orientations only in a narrow middle region according to hyperbolic tangent functions, once in an increasing, then in a decreasing manner. In both of these cases $\phi(x)$ was set to a value slightly below 1 (to avoid the singularity of $g(\phi)$ at $\phi=1$), with a small dip added in the middle region. Convergence of the solutions was checked by monitoring the decrease of the total free energy of the system via Eq.~\ref{eq:originalf} and~\ref{eq:foriHMP}.

Figure~\ref{fig:GBProfile13} shows the two different solutions that we obtained after long enough simulation time, when no further decrease of the free energy could be observed. The two different $\theta(x)$ profiles map to opposite segments of the order parameter space. The solution corresponding to the shorter path ($\Delta\theta=0.8\pi$) has a smaller dip in $\phi(x)$ and lower total free energy than the solution corresponding to the longer path ($\Delta\theta=1.2\pi$). The two solutions cannot continuously transform to one another.

\subsubsection{Results of the US model}

The same setup as above was simulated with the unit sphere model. The results were obtained by solving Eq.~\ref{eq:phidot2} and~\ref{eq:thetadot2} with boundary conditions $\phi^\prime(x_A)=0$, $\boldsymbol{\theta}(x_A)=(\cos(\theta_A),\sin(\theta_A),0)$ on the left and $\phi^\prime(x_B)=0$, $\boldsymbol{\theta}(x_B)=(\cos(\theta_B),\sin(\theta_B),0)$ on the right ends. The initial conditions were the same $\phi(x)$ and $\boldsymbol{\theta}(x)=(\cos(\theta(x)),\sin(\theta(x)),0)$ with the same $\theta(x)$ as used with the HMP model.

\begin{figure}
\centerline{\includegraphics[height=0.5\linewidth]{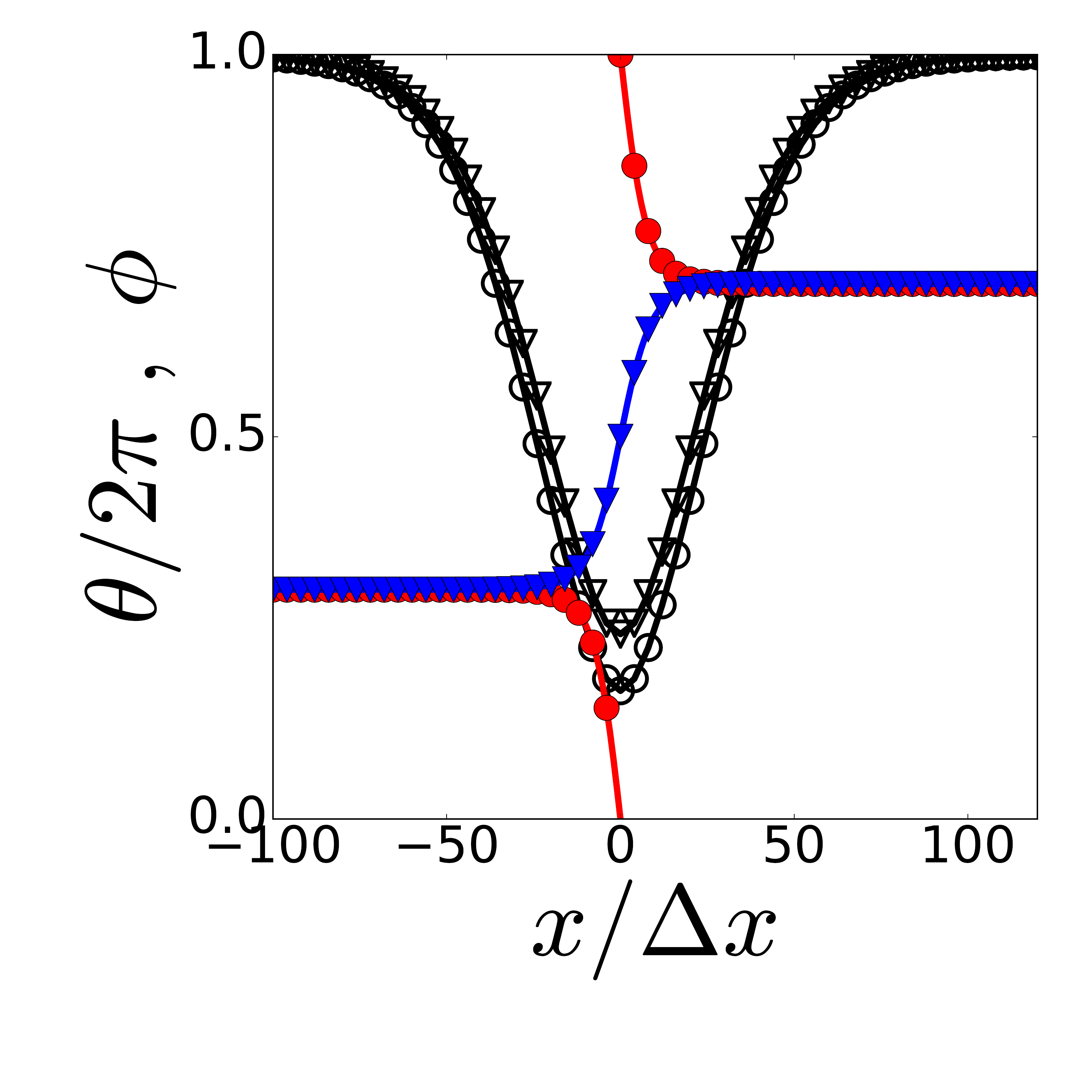}\includegraphics[height=0.5\linewidth]{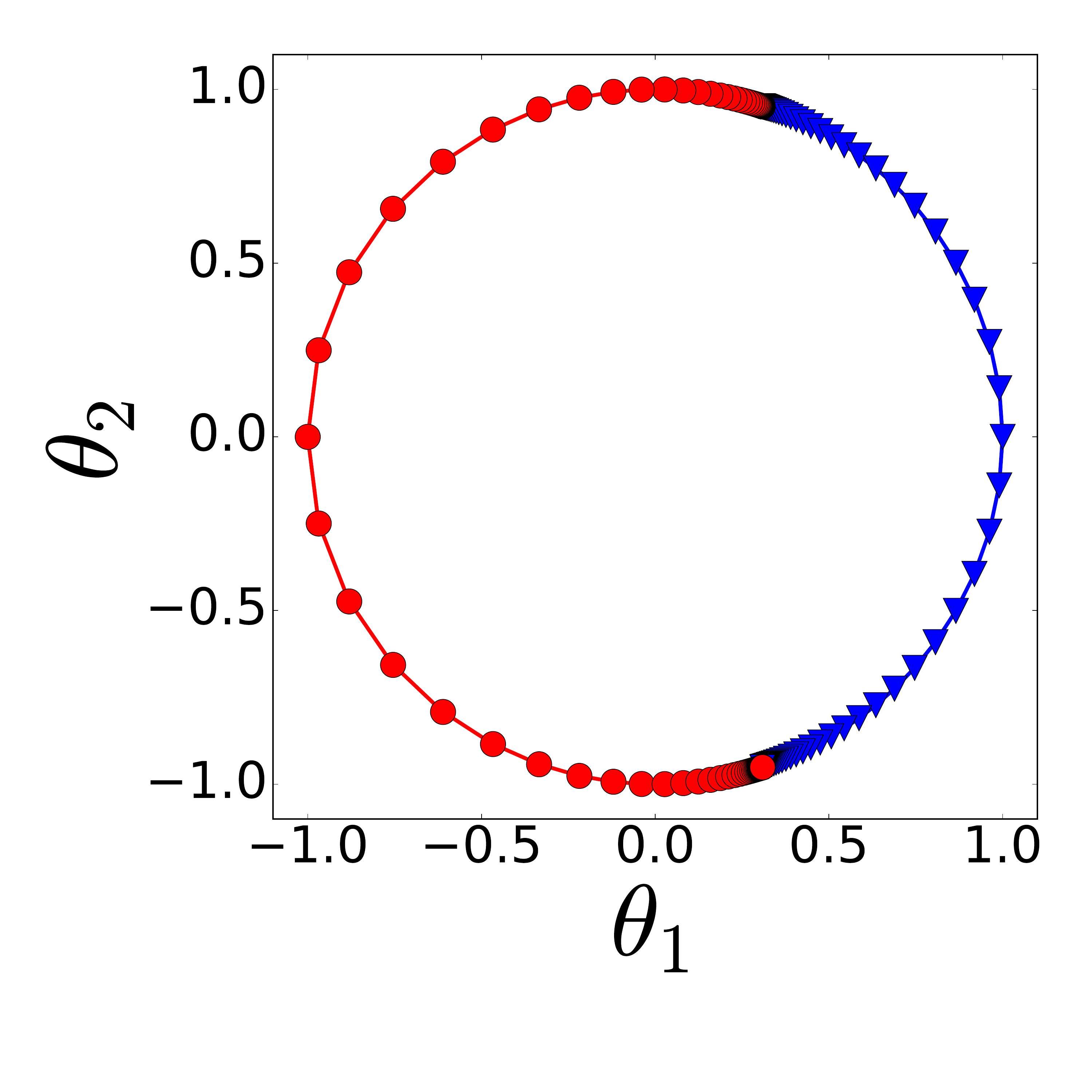}}
\begin{tabularx}{\linewidth}{XX}
\centering (a) & \centering (b)
\end{tabularx}
\caption{The two different grain boundary solutions of the HMP model (solid lines) and the respective two solutions of the US model (symbols). For the latter, $\theta_3$ remained zero, and the scalar $\theta$ plotted is the polar angle of the vector $(\theta_1,\theta_2)$. The results produced by the two models are visually indistinguishable. The triangles/circles and the continuous lines underneath them correspond to the lower/higher energy solutions. The open black symbols show to the phase field, the closed red and blue symbols show the orientation field. (a) profiles in the real space, (b) profiles in the order parameter space. Neighboring symbols correspond to neighboring cells in the simulation domain.}
\label{fig:GBProfile13}
\end{figure}

The equilibrium profiles obtained are plotted on top of the respective profiles of the HMP model in Figure~\ref{fig:GBProfile13}. As initially $\theta_3$ was set to zero, it remained zero throughout the simulation, as expected. For the 3-component model, the actual orientation $\theta$ is defined via the relations $\theta_1=\cos(\theta)$ and $\theta_2=\sin(\theta)$. The solutions of the two models are indistinguishable, in agreement with our previous statement, that in the $\theta_3=0$ and small $\Delta x$ limit the two models are equivalent.

\begin{figure}
\centerline{\includegraphics[width=\linewidth]{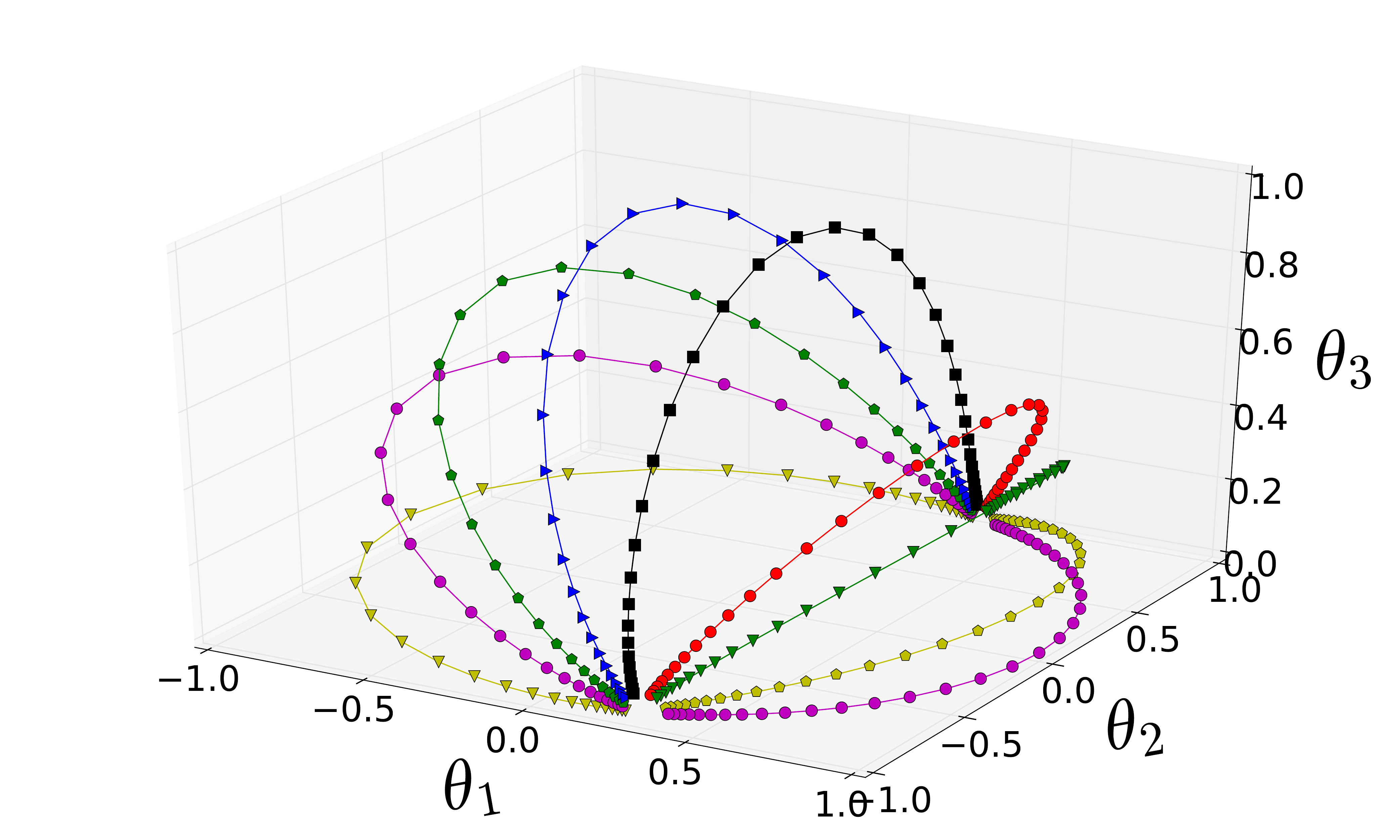}}
\caption{The transition from the unstable higher energy grain boundary solution to the stable lower energy grain boundary solution in the unit sphere model. The 9 lines shown correspond to grain boundary profiles at different simulation times, mapped to the order parameter space. Time is increasing from left to right. To trigger the transition, a small value was added to $\theta_3$ in the beginning of the simulation.}
\label{fig:GBTransition3}
\end{figure}

Next, we checked the stability of the two solutions. We took the equilibrium profiles just obtained and added a small value ($10^{-3}$) to $\theta_3$, and renormalized $\boldsymbol\theta$ to $|\boldsymbol\theta| = 1$. Then we started new simulations with these slightly modified profiles as initial conditions. In case of the lower energy profiles (black and blue triangles in Fig.~\ref{fig:GBProfile13}), the system relaxed back to the initial profiles with $\theta_3=0$, indicating that this solution corresponds to a stable equilibrium. In contrast, when we started from the higher energy profiles (black and red circles in Fig.~\ref{fig:GBProfile13}), the system did not relax back to the original solution, instead, it transformed to the lower energy one~(Fig.~\ref{fig:GBTransition3}). This indicates that the higher energy profile corresponds to an unstable equilibrium.

These simulations show us that the topologically stable higher energy solutions of the HMP model become topologically and energetically unstable solutions of the US model. Therefore, by adding small perturbations to $\theta_3$, the system can relax to the lower energy grain boundary solutions via intermediate $\theta_3 \ne 0$ states.

\subsubsection{Results of the LDG model}
\label{sec:1dldg}

\begin{figure}
\centerline{\includegraphics[height=0.5\linewidth]{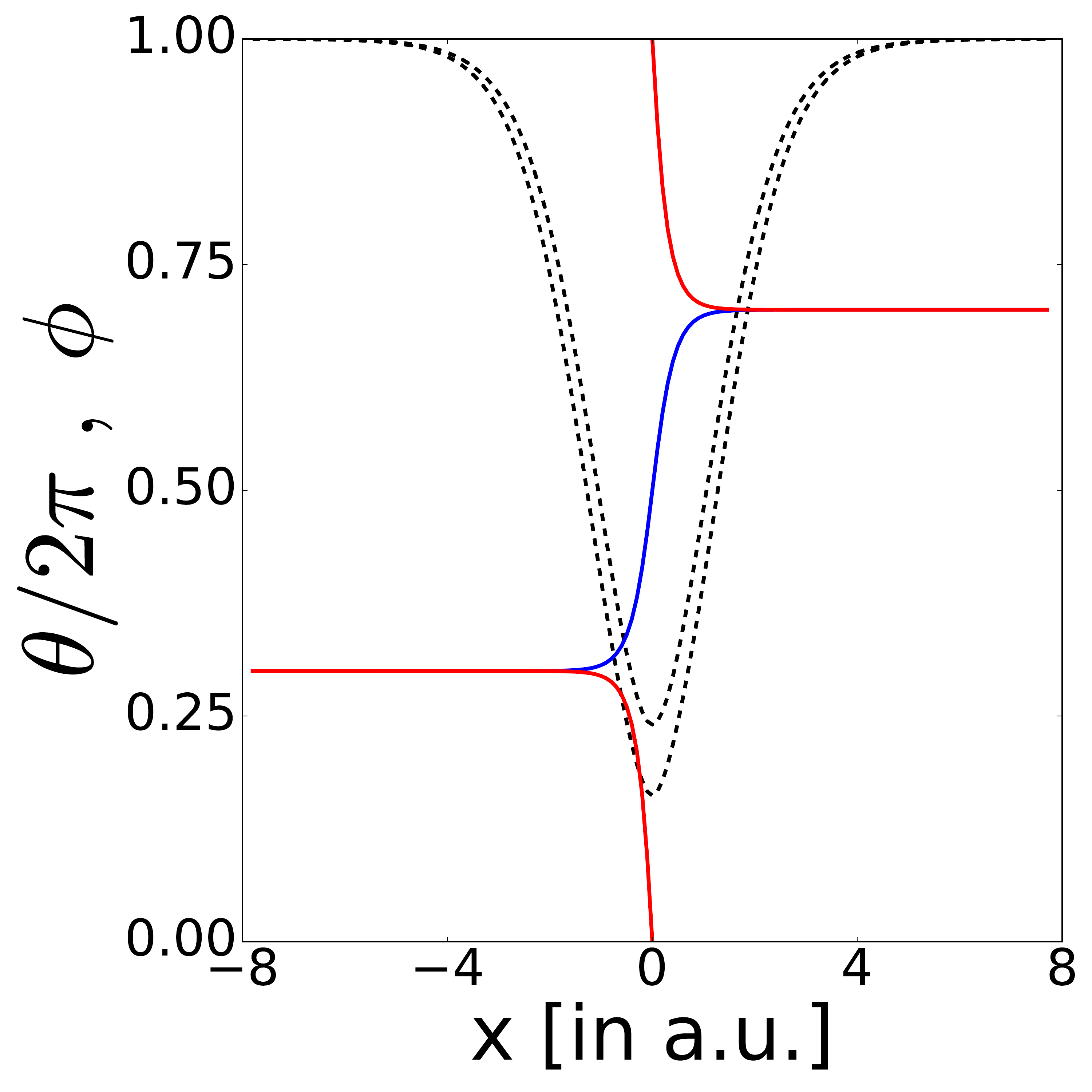}\raisebox{-0.01\linewidth}{\includegraphics[height=0.5\linewidth]{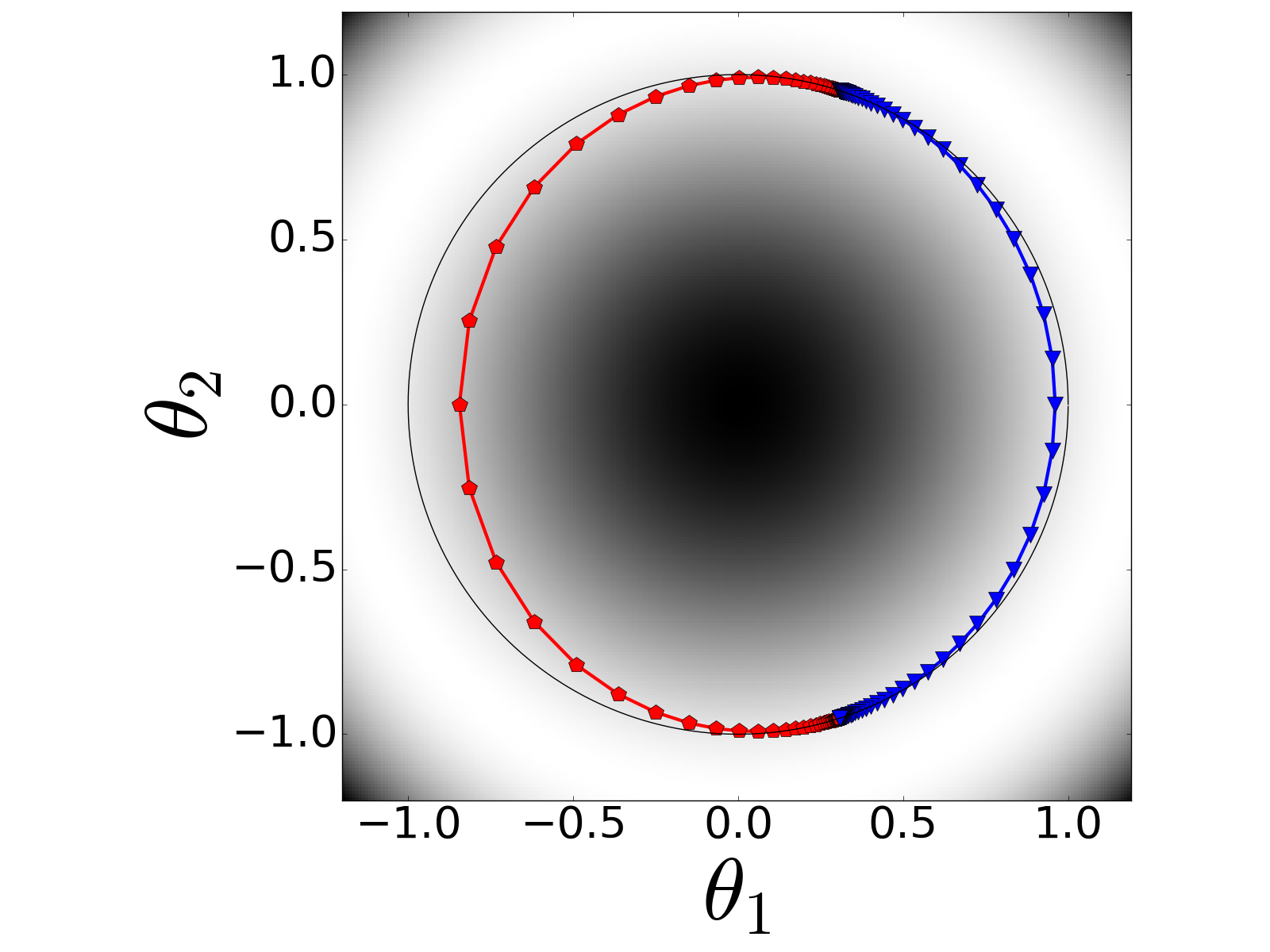}}}
\begin{tabularx}{\linewidth}{XX}
\centering (a) & \centering (b)
\end{tabularx}
\caption{Grain boundary solutions obtained by the LDG model for $\nu=50$. With this value of $\nu$ the barrier is high enough for two separate solutions to exist. (a) the $\phi$ and $\theta$ profiles in real space, (b) the $\boldsymbol{\theta}$ profiles in the order parameter space. The background shading is according to $f_s(\boldsymbol\theta)$: white corresponds to low, black corresponds to high values of the potential. The black solid line is the unit circle.}
\label{fig:GBProfile4_1}
\end{figure}

\begin{figure}
\centerline{\includegraphics[height=0.47\linewidth]{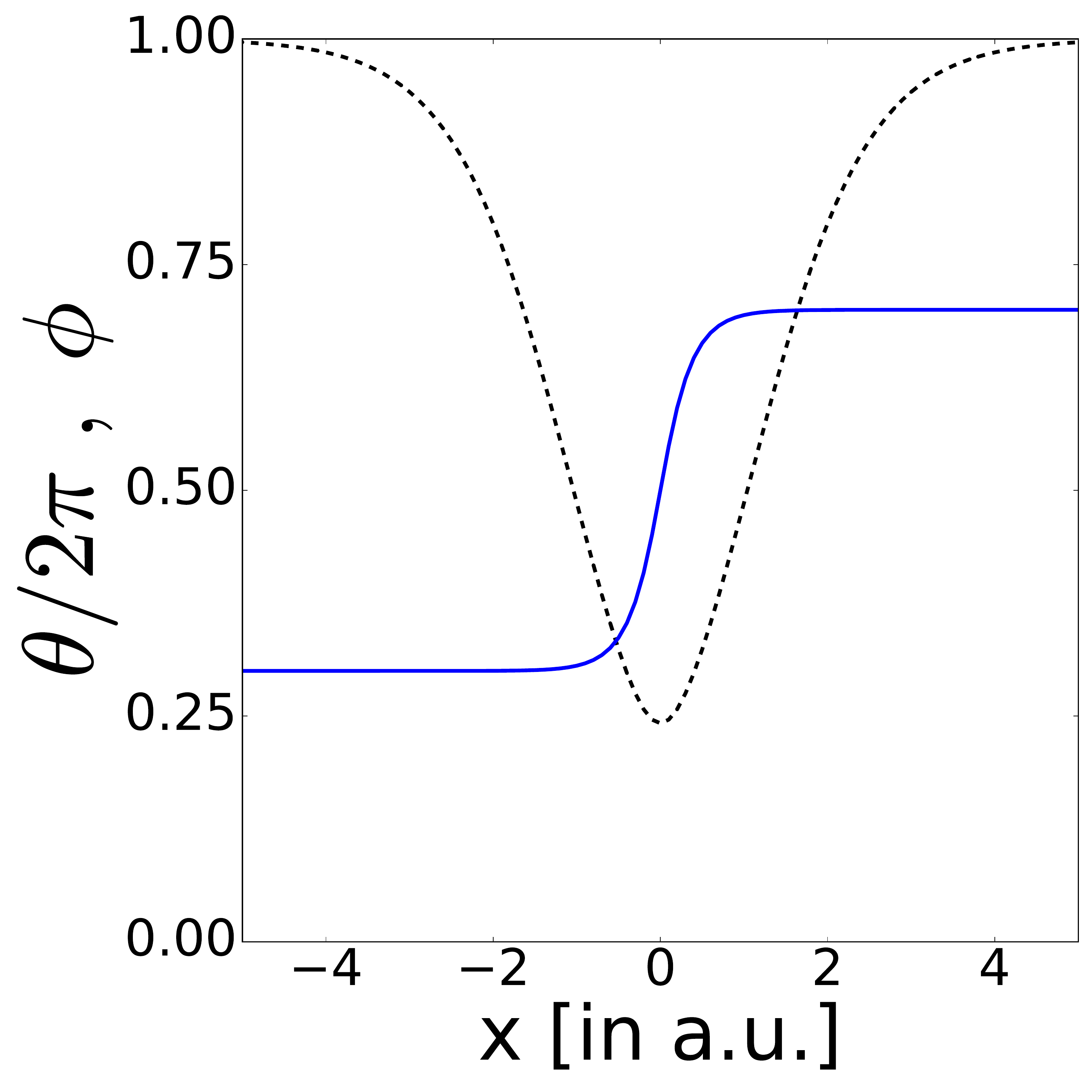}\raisebox{-0.03\linewidth}{\includegraphics[height=0.51\linewidth]{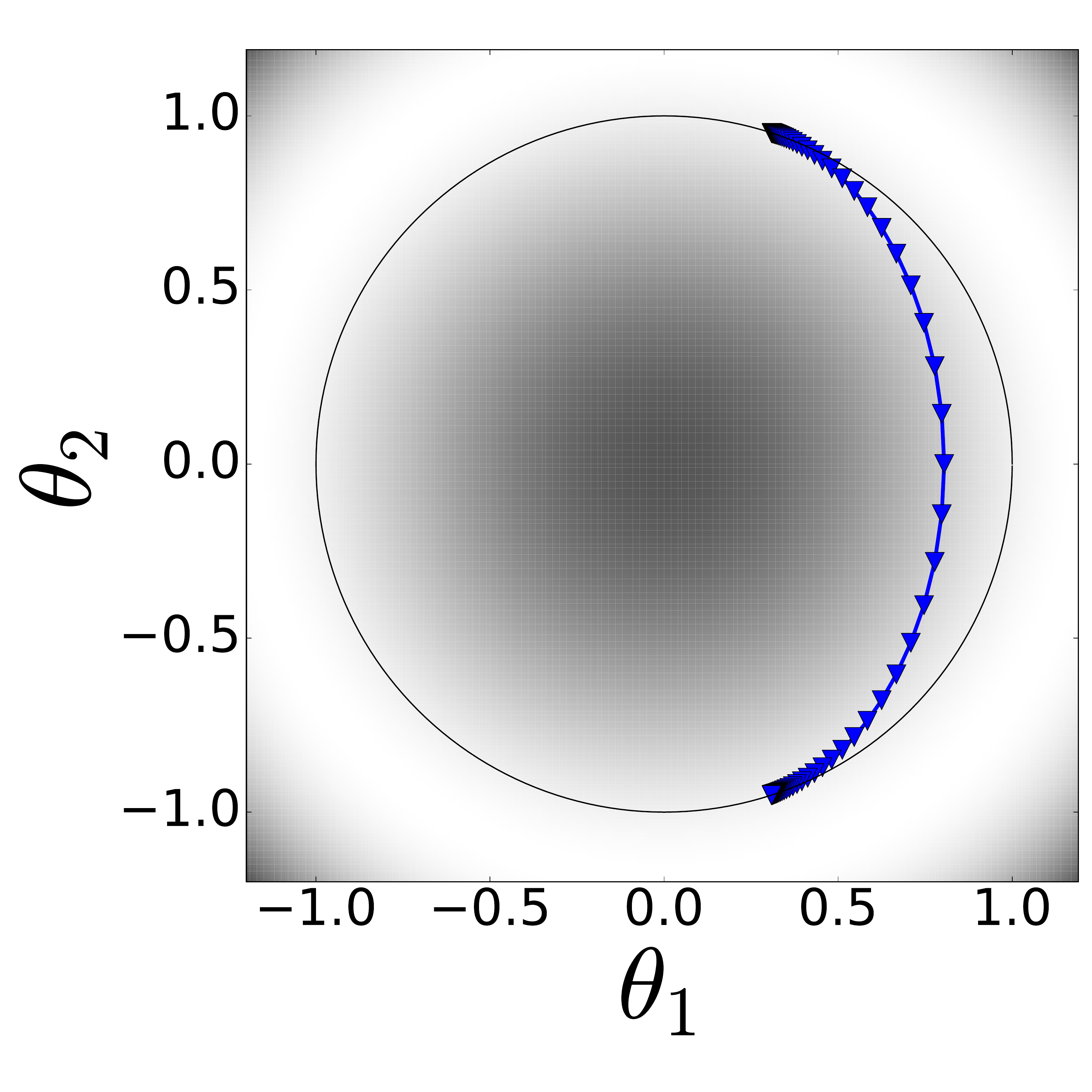}}}
\begin{tabularx}{\linewidth}{XX}
\centering (a) & \centering (b)
\end{tabularx}
\caption{The grain boundary solution obtained by the LDG model for $\nu=10$. With this value of $\nu$ the barrier is low and only one solution, the one corresponding to the lower energy solution in Fig.~\ref{fig:GBProfile4_1} exists. For further description of the figure elements, see the caption of Figure~\ref{fig:GBProfile4_1}.}
\label{fig:GBProfile4_2}
\end{figure}

\begin{figure}
\centerline{\includegraphics[height=0.5\linewidth]{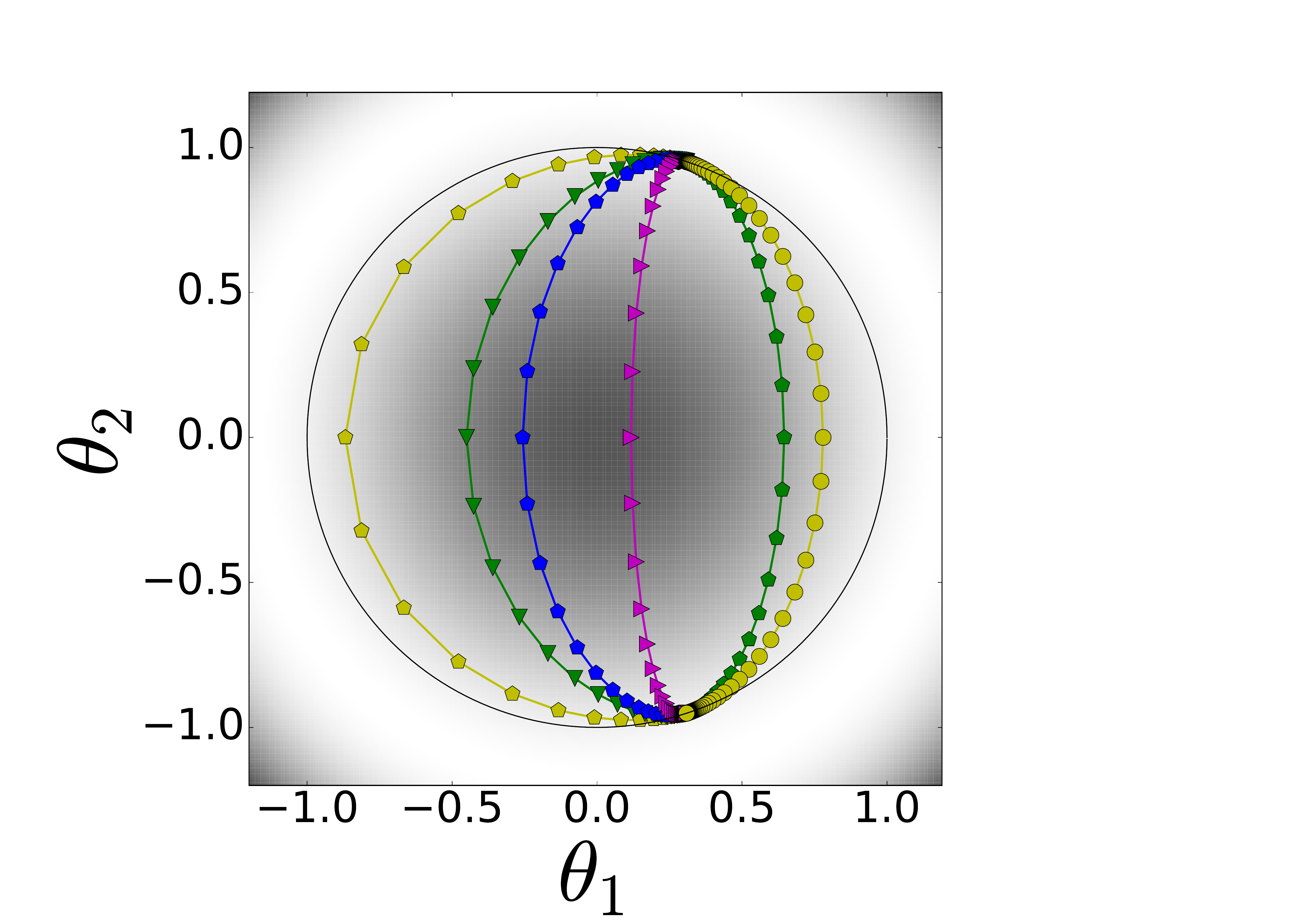}}
\caption{The transition of the $\boldsymbol{\theta}$ profile from the profile corresponding to the higher energy solution of the HMP model to the solution of the 2-component vector field model with $\nu = 10$, shown in the order parameter space. As in Figure~\ref{fig:GBProfile13}, time increases from left to right.}
\label{fig:GBProfile4_3}
\end{figure}

Finally, we repeated the same procedure with the LDG model. With the exception of the unneeded $\theta_3$ component, we used the same boundary and initial conditions as with the US model. Depending on the value of $\nu$ which sets the magnitude of the potential $f_s({\boldsymbol\theta})$, different behavior of the model is observed. If $\nu$ is large ($\nu>\nu_\mathrm{crit}$), then the potential has a high local maximum at the origin and a steep valley along its minimum that follows the unit circle in the $\theta_1,\theta_2$ plane. In this case the two different initial conditions result in different equilibrium solutions (see Fig.~\ref{fig:GBProfile4_1}) that are separated by the high peak of the potential in the centre. Both solutions are stable, but in contrast to the HMP model, this stability is not topological, it results from the high energy barrier between the two paths. If $\nu$ is small ($\nu<\nu_\mathrm{crit}$), however, the height of the potential is not sufficient  to separate the two solutions. In this case, only one solution exists (see Fig.~\ref{fig:GBProfile4_2}). The value of $\nu_\mathrm{crit}$ increases with increasing $\epsilon$. The transition of the high energy profile to the lower energy solution is shown in Fig.~\ref{fig:GBProfile4_3}.

\subsection{Structure of the defects}

In this subsection the structure of the orientation field along a grain boundary with an isolated topological defect is investigated using the different models. We chose a symmetric setup which corresponds to a grain boundary with misorientation $\Delta\theta = \pm\pi$, as in this (and only in this) symmetric case there is no driving force for the defect to move. To construct appropriate initial conditions for the model investigated, we first determined the two different 1D equilibrium grain profiles corresponding to $\Delta\theta = \pi$ (using the same method shown in the previous subsection) with the respective model, and then we made a ``hybrid'' grain boundary in 2D by placing these different profiles in the left and right halves of the simulation domain. The final equilibrium profiles are then obtained as the fully relaxed long-time solutions of the respective governing equations. 

If not stated otherwise, we used a 2D domain of $256 \times 256$ pixels with a grid size of $h=0.05$ in all simulations below. For the phase-field we applied Neumann boundary conditions with zero normal derivatives on all sides. For the orientation fields we used mixed boundary conditions, the details will be given later in the model specific descriptions.

\subsubsection{Defect structure in the HMP model}

\begin{figure}
\centerline{\includegraphics[width=0.33\linewidth]{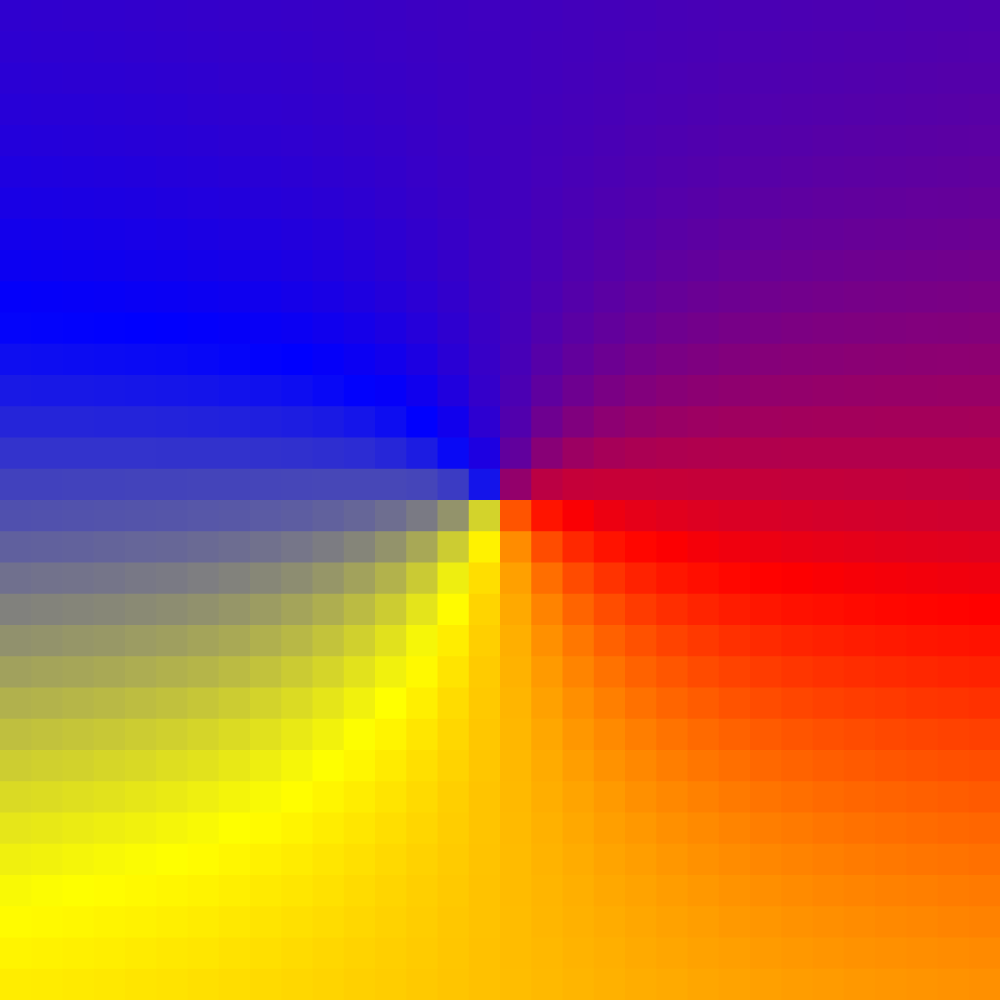}~\includegraphics[width=0.33\linewidth]{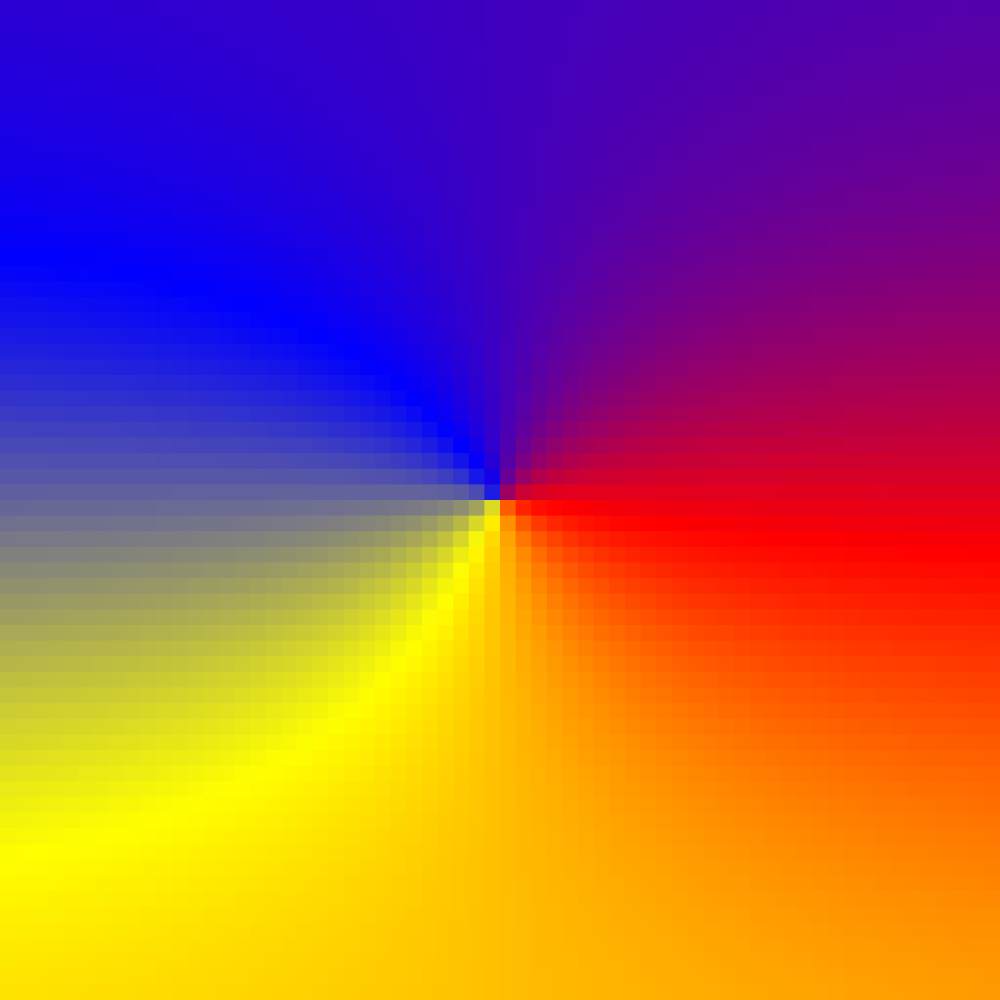}~\includegraphics[width=0.33\linewidth]{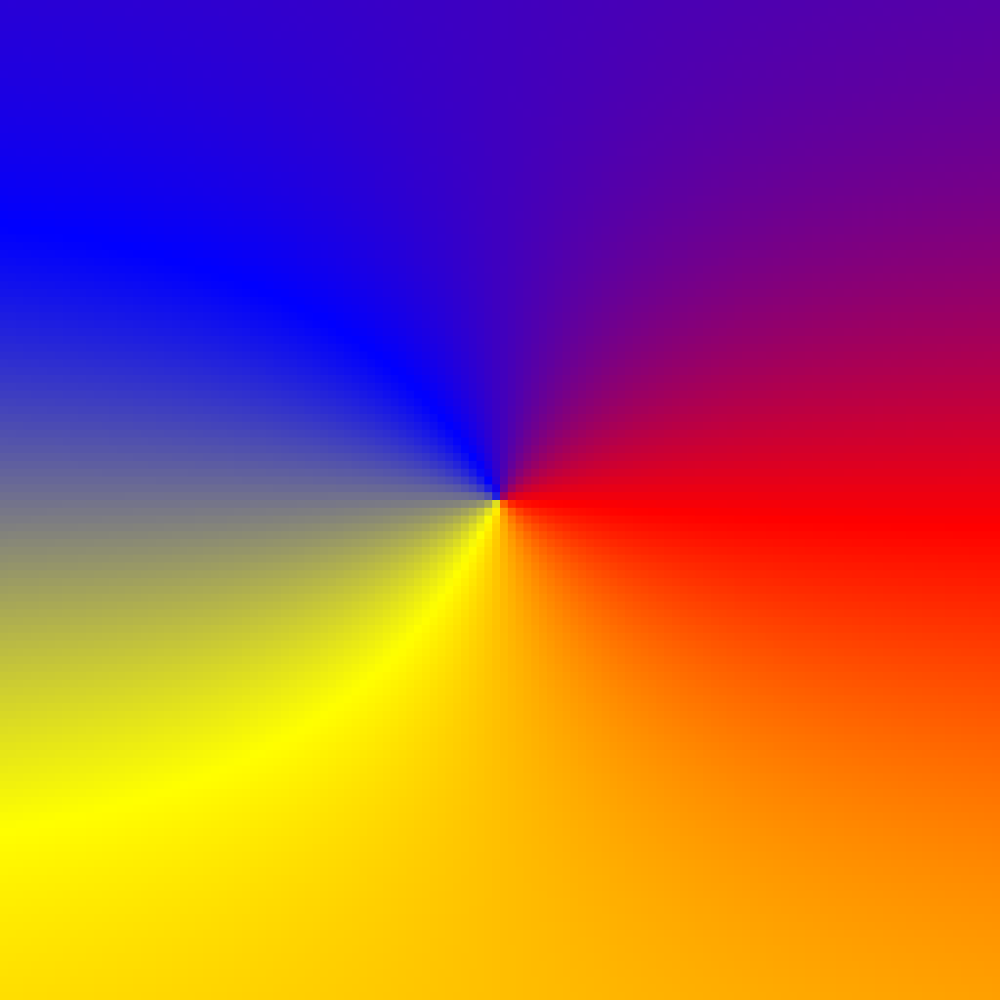}}
\centerline{\includegraphics[width=0.33\linewidth]{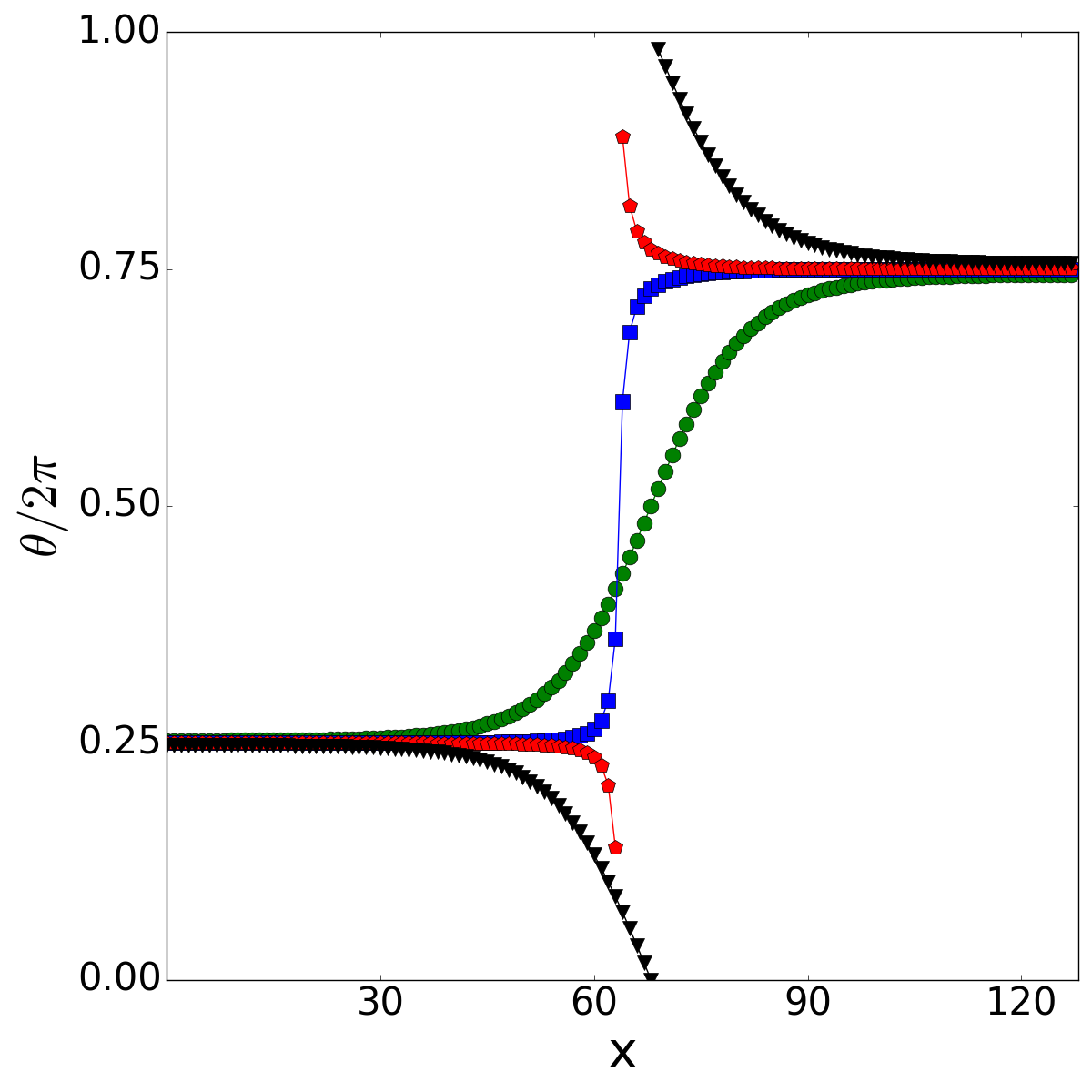}~\includegraphics[width=0.33\linewidth]{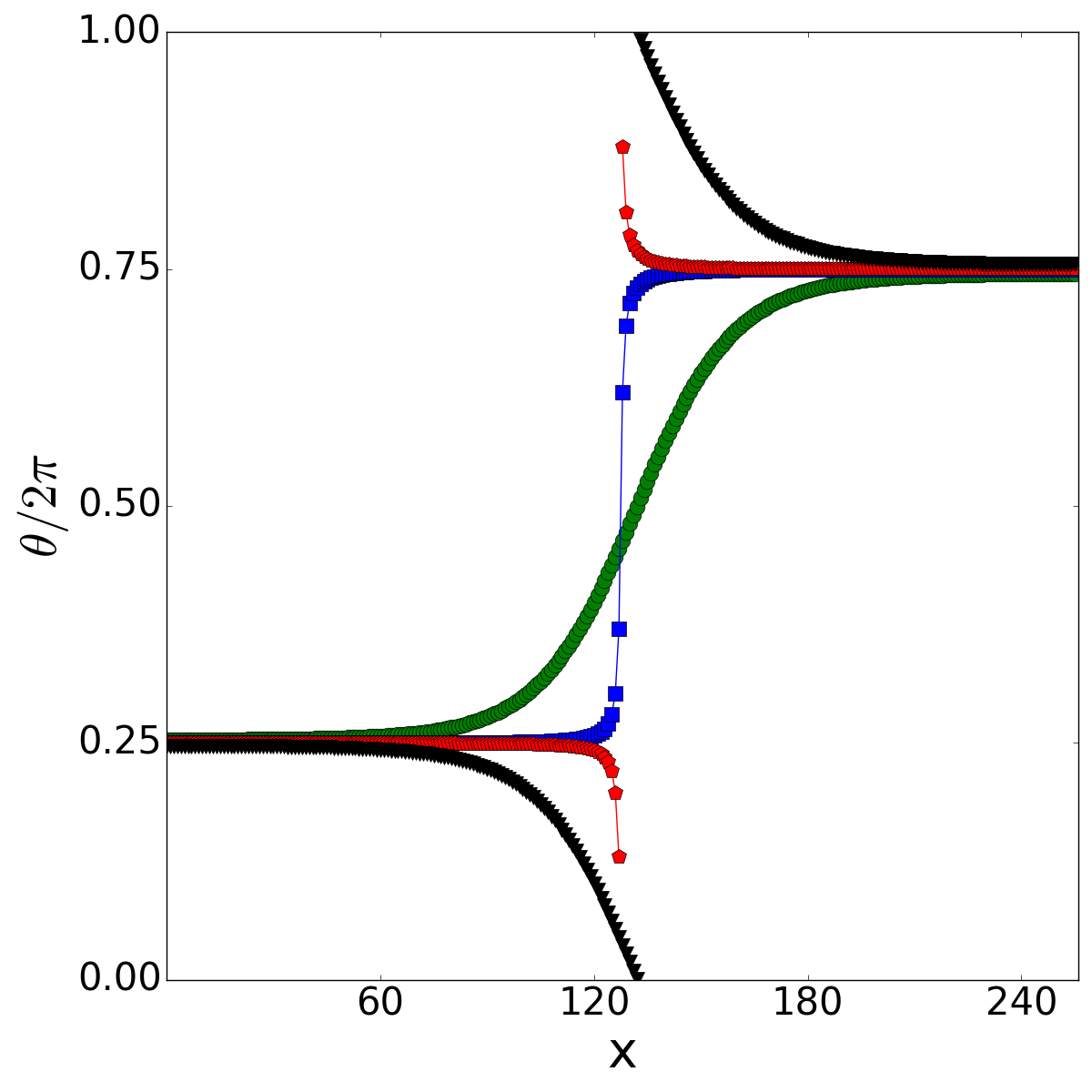}~\includegraphics[width=0.33\linewidth]{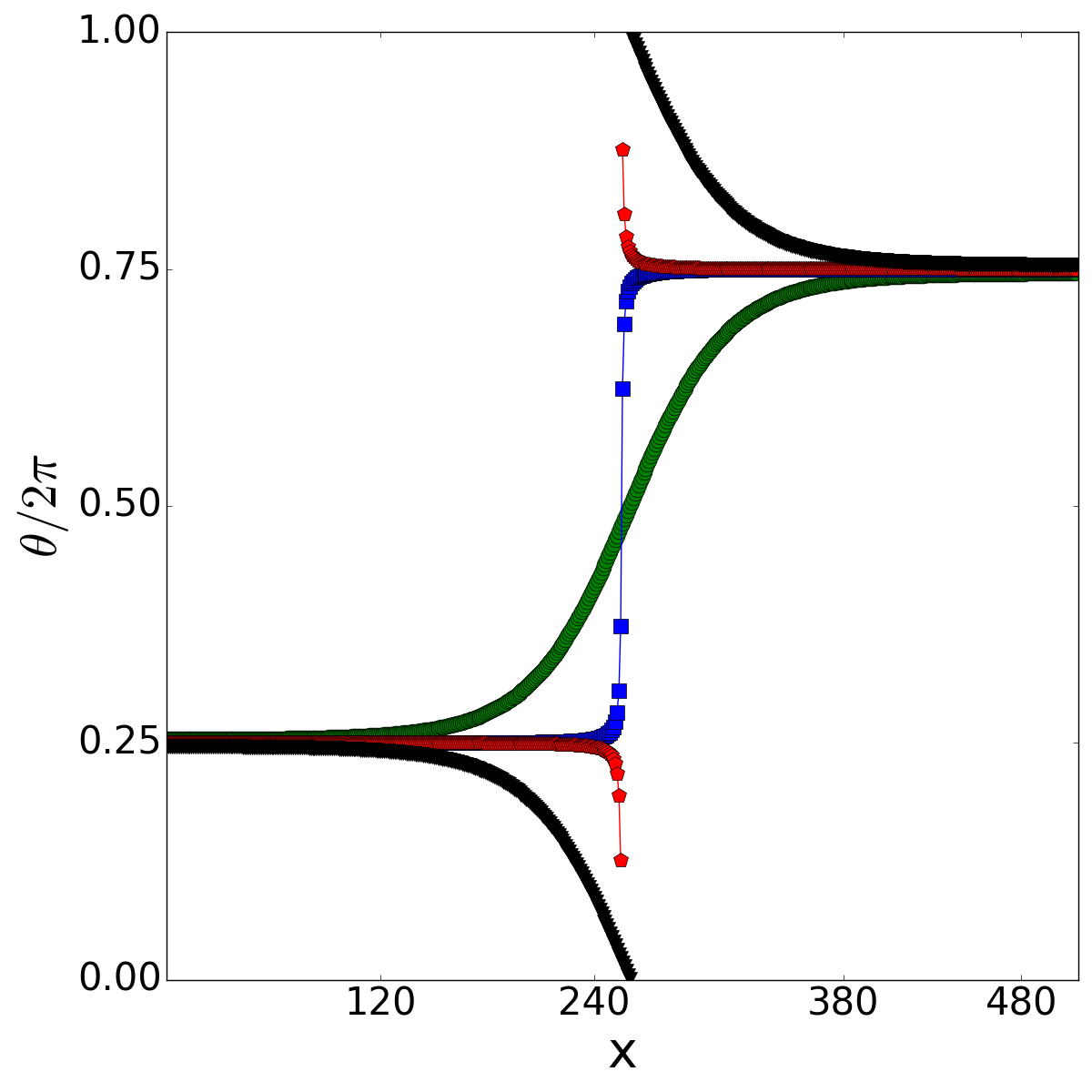}}
\begin{tabularx}{\linewidth}{XXX}
\centering (a) & \centering (b) & \centering (c)
\end{tabularx}
\caption{Structure of the orientation field around the equilibrium defect in the HMP model with $\Delta x = 0.05$ (a) $0.025$ (b) and $0.0125$ (c). The top row shows the color map of $\theta$ in a small region around the defect, strongly enlarged for better visibility. We used a circular color map which assigns the colours red, blue, yellow and again red to the orientations $\theta = 0, 2\pi/3, 4\pi/3, 2\pi$, and interpolates the RGB values of these colors for orientations in between. The bottom row shows the respective 1D profiles. The four lines shown correspond to the leftmost (black triangles), rightmost (green circles) and the two middle (blue squares and red pentagons) columns of the simulation domain.}
\label{fig:defect1c}
\end{figure}

To fix the bulk grain orientations but allow for a smooth transition between them across the grain boundary we applied Dirichlet boundary conditions $\theta=\pi/2$ on the top and $\theta=3\pi/2$ on the bottom sides and Neumann boundary conditions with zero normal derivatives on the left and right sides of the domain. The first column of Figure~\ref{fig:defect1c} shows the equilibrium structure obtained by simulating Eq.~\ref{eq:phidot0} and~\ref{eq:thetadot0}. To illustrate the effect of the grid resolution on the result, we repeated the simulation using two finer meshes with $h=0.025$ and $h=0.0125$. The results are shown in the remaining two columns of Figure~\ref{fig:defect1c}. Please notice that the orientational difference between neighboring cells very close to the defect (shown by the well separated blue squares and red pentagons in the bottom line) is independent of the grid resolution used, meaning that it is only the grid which limits the gradient of the orientation field. This is in agreement with the expected singularity of the orientation field at this point.

\subsubsection{Defect structure in the US model}

For this model too, the boundary conditions were chosen to correspond to the boundary conditions used in the HMP model. On the top and bottom sides we used Dirichlet boundary conditions with $\boldsymbol\theta = (1,0,0)$ and $\boldsymbol\theta = (-1,0,0)$, while on the left and right sides we used Neumann boundary conditions with zero normal derivatives for $\theta_1$ and $\theta_2$, but fixed the value of $\theta_3$ at $\theta_3 = 0$.

\begin{figure}
\setlength{\tabcolsep}{0pt}
\begin{tabular}{m{0.35\linewidth}m{0.65\linewidth}}
\includegraphics[width=\linewidth]{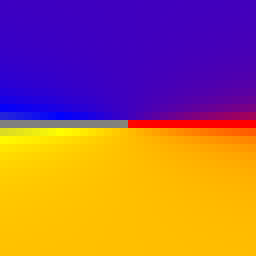}&\includegraphics[width=\linewidth]{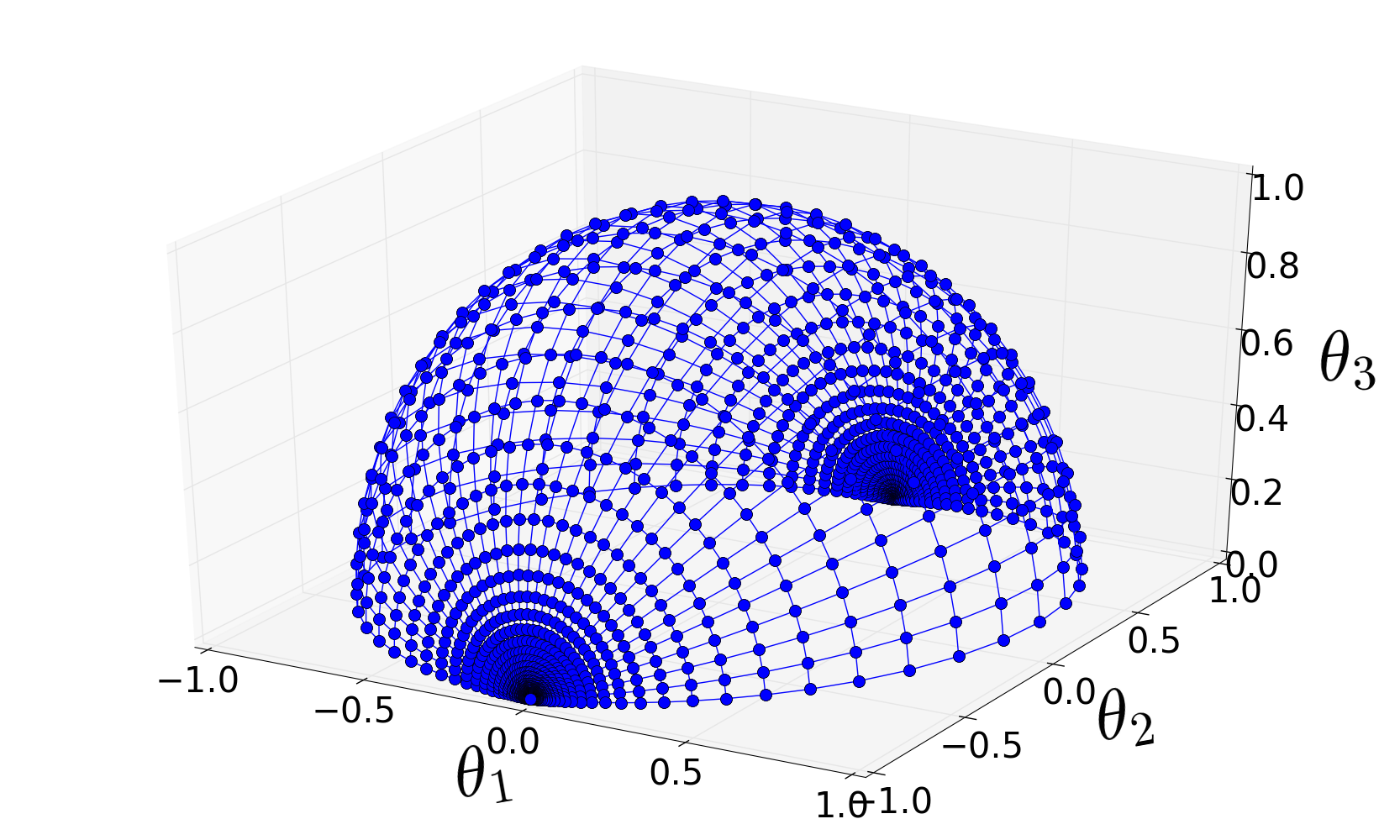}
\end{tabular}
\begin{tabularx}{\linewidth}{XXXXXXXX}
~ & \centering (a) & ~ & ~ & ~ &\centering (b) & ~ &
\end{tabularx}
\caption{Structure of the orientation field around the equilibrium defect in the unit sphere model. (a) color map of the scalar orientation field $\theta$. For further explanation, see the caption of Figure~\ref{fig:defect1c}. (b) the order parameter space map of the vector orientation field $\boldsymbol{\theta}$. Points corresponding to neighboring pixels are shown connected.}
\label{fig:defect3c}
\end{figure}

The equilibrium structure obtained by simulating Eq.~\ref{eq:phidot1} and~\ref{eq:thetadot1} is shown on Figure~\ref{fig:defect3c}. The right panel shows the solution mapped to the order parameter space. The points that correspond to neighboring pixels of the simulation are connected with lines. The points $(\pm1,0,0)$ correspond to the bulk orientations along the top and bottom sides, while the points on the left and right half circles in the $\theta_3=0$ base plane correspond to the grain boundary profiles along the left and right sides of the simulation domain. The fine mesh structure formed by the connection lines shows that the vector order parameter field $\boldsymbol\theta$ is continuous even in the neighborhood of the defect. This is not true for the corresponding scalar orientation field $\theta$. A singularity similar to the one seen in the HMP model is visible at the centre.

\subsubsection{Defect structure in the LDG model}

\begin{figure}
\setlength{\tabcolsep}{0pt}
\begin{tabular}{m{0.4\linewidth}m{0.6\linewidth}}
\raisebox{0.2in}{\includegraphics[width=\linewidth]{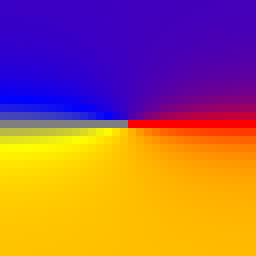}}&\includegraphics[width=0.85\linewidth]{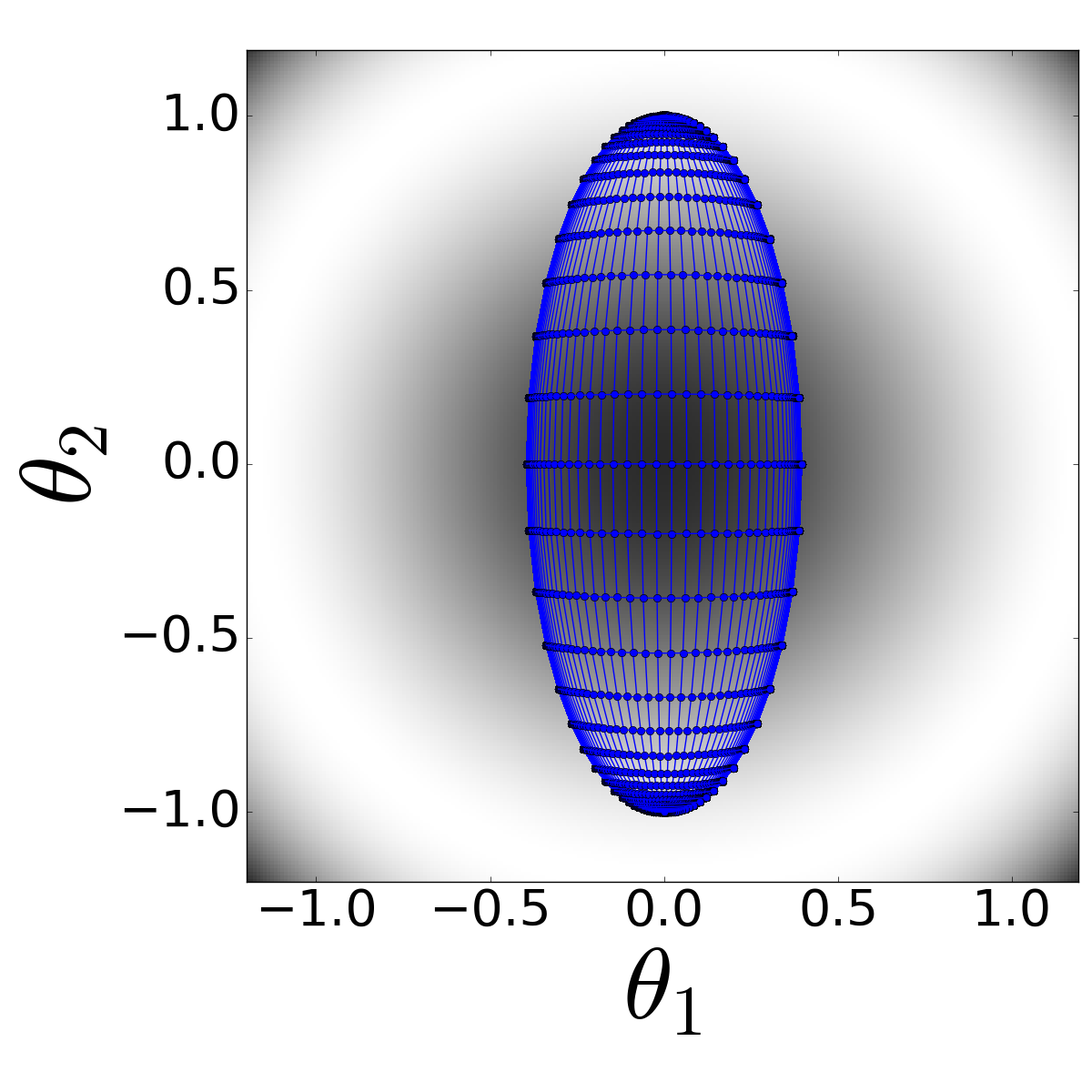}
\end{tabular}
\begin{tabularx}{\linewidth}{XX}
\centering (a) & \centering (b)
\end{tabularx}
\caption{Structure of the orientation field around the equilibrium defect in the LDG model with $\nu=10$, when two stable grain boundary solutions and therefore a stable defect exists. (a) color map of the scalar orientation field $\theta$, (b) the order parameter space map of the vector orientation field $\boldsymbol{\theta}$. For further explanation, see the captions of Figure~\ref{fig:defect1c} and~\ref{fig:defect3c}.}
\label{fig:defect2c}
\end{figure}

The boundary conditions of this model are the same as the boundary conditions of the first two components of the 3-component unit vector field model. We simulated Eq.~\ref{eq:phidot2} and~\ref{eq:thetadot2} with $\nu = 10$. The results are shown in Figure~\ref{fig:defect2c} in a similar way as in the case of the unit sphere model. As we can see in the figure, similar to the US model, the vector orientation field is smooth, while the derived scalar orientation field is not.

The present choice of $\nu = 10$ provides a barrier high enough to separate the two solutions sufficiently in the symmetric case studied here, but the same value was not high enough to keep the higher energy solution stable for the slightly asymmetric case shown in Figure~\ref{fig:GBProfile4_3}. If we started to decrease $\nu$, the ellipsoidal shape of the solution in the order parameter space (see Fig.~\ref{fig:defect2c}) would get thinner, collapsing finally to a straight line. This would mean no further ambiguity in the grain boundary solutions and therefore the disappearance of the defect.

\subsection{Elimination of lattice pinning}
\label{sec:pinning}

As we have just illustrated, the existence of two different continuous grain boundary solutions result in topological defects in the HMP model if the two different profiles co-exist on the same grain boundary. The singularity corresponding to these defects may cause problems in numerical simulations. The most important one we observed is the pinning of these defects by the simulation grid. First, we exemplify the phenomena using the original HMP model, then we show that the newly proposed models are free of this problem. 

For these simulations, we took the same setup as we used in the previous section to start with. We introduced a small driving force for the defect to move by making the setup slightly asymmetric. We achieved this by changing the boundary conditions for the orientation field on the top and bottom boundaries to correspond to $\theta_A = \pi/2 + \epsilon$ and $\theta_B = 3\pi/2 - \epsilon$ with $\epsilon=\pi/5$. In this setup, a drift of the defect toward the high energy grain boundary is expected and its velocity should be proportional to $\epsilon$, i.e., to the gain of energy induced by the drift. To make room for the defect to move, we extended the sample in the direction of the grain boundary by using a domain of $2048 \times 256$ pixels with $\Delta x = 0.05$. Generally, in a numerical simulation with proper discretization (both in space and time) we should observe this drift, and by increasing the resolution further and further, we should see a convergence of the drift velocity towards its limiting value that corresponds to the continuous case.

\subsubsection{Pinning in the HMP model}
\label{sec:pinningHMP}

\begin{figure}
\includegraphics[width=0.5\linewidth]{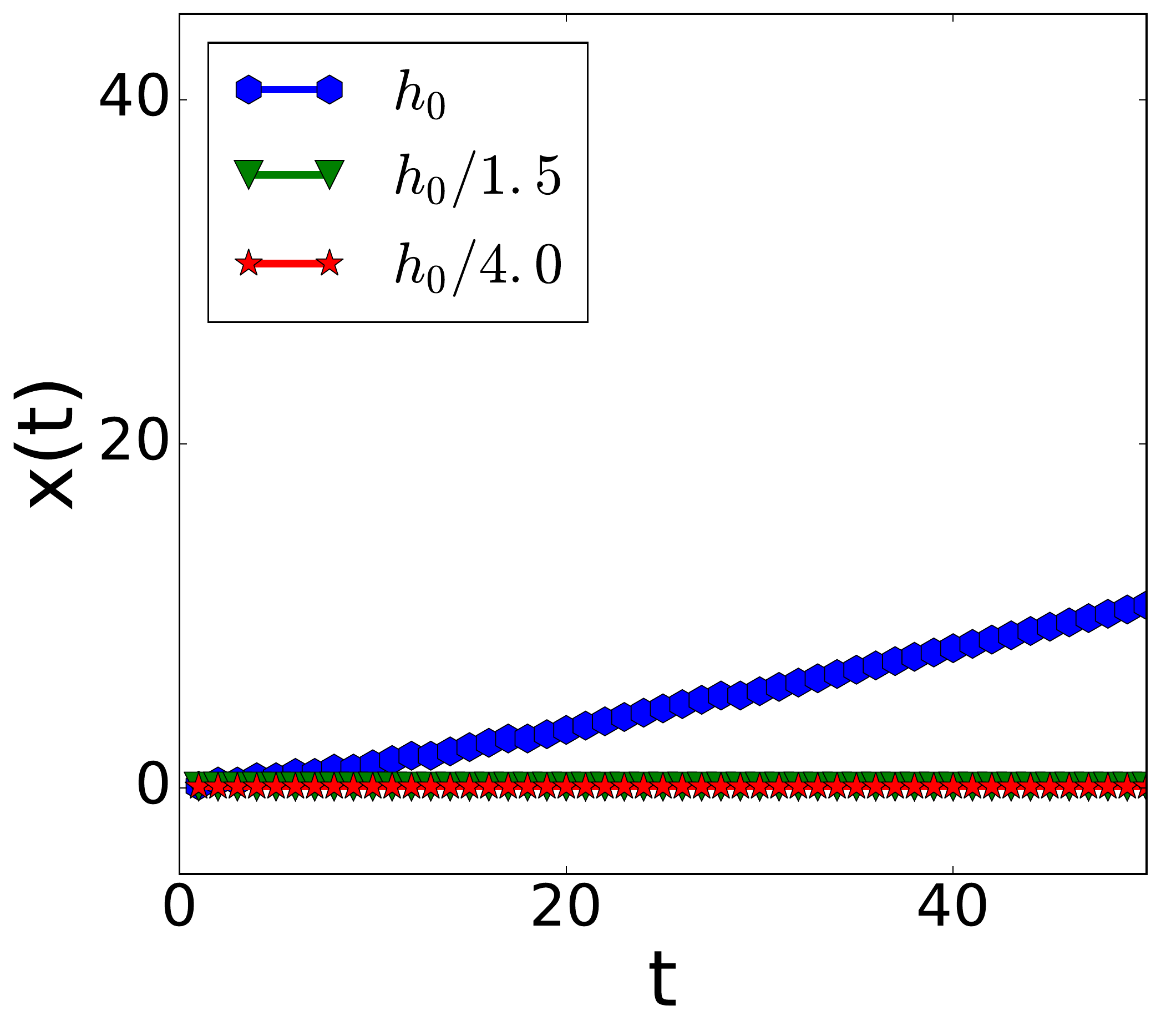}\includegraphics[width=0.5\linewidth]{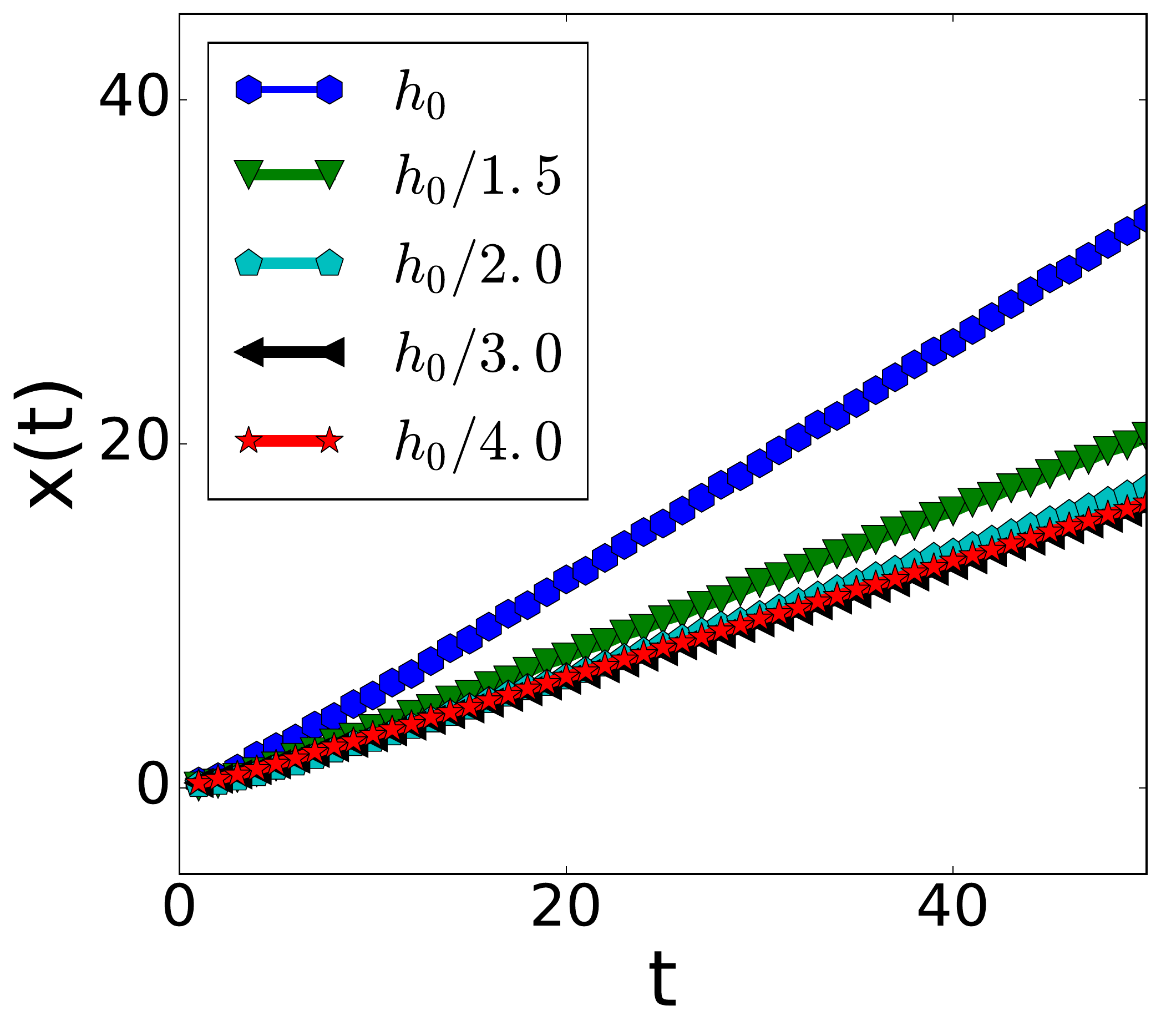}
\begin{tabularx}{\linewidth}{XX}
\centering (a) & \centering (b)
\end{tabularx}
\caption{Defect position vs.~time in an asymmetric setup in the HMP (a) and LDG (b) models. Time and space coordinates are in arbitrary units. The reference grid spacing is $\Delta x_0 = 0.05$. In case of the HMP model we see the pinning of the defect as the spatial resolution increased, while in case of the LDG model a convergence to a constant drift velocity is observed.}
\label{fig:pinning}
\end{figure}

As in Figure~\ref{fig:defect1c}, we simulated this setup with three different grid resolutions. We determined the location of the defect and plotted its position vs.~time in Figure~\ref{fig:pinning}, left. At the lowest resolution we observed that the defect started to drift, building up a constant velocity. This is the expected behavior, as it makes the length of the low energy grain boundary increase at the expense of the high energy one, thus decreasing the total free energy of the system. Surprisingly, when we used finer and finer grid resolutions to simulate the same setup, instead of a convergence of the drift velocity to a finite limiting value, we observed that the defect got stuck in its original position. This pinning is certainly a numerical issue and related to the fact that the defect is singular. In a discretized non-singular 2D system $\nabla\theta$ does not change when we decrease the grid spacing $\Delta x$, which means that the contribution of a pixel to the discrete free energy scales with the grid spacing as $\Delta x^2$. At the singularity, however, not $\nabla\theta$, but the difference $\Delta\theta$ between neighboring pixels remains independent of $\Delta x$ (see Figure~\ref{fig:defect1c}), meaning that the gradient scales as $1/\Delta x$. Thus, if there is a $|\nabla\theta|^2$ term in the free energy density, the contribution of such pixel to the discrete free energy is independent of $\Delta x$ . As a result, when decreasing $\Delta x$ the singular region is becoming dominant and prevents motion.


\subsubsection{No pinning in the US and LDG models}

We have shown that in the US model and also in the LDG model with $\nu<\nu_\mathrm{crit}(\epsilon)$, grain boundaries with higher energy will relax spontaneously toward the low energy ones, therefore, such defects do not exist and therefore the issue of lattice pinning is irrelevant. The remaining case is the LDG model with $\nu>\nu_\mathrm{crit}(\epsilon)$ where both grain boundary solutions are stable (see Section~\ref{sec:1dldg}). Simulating this case with different spatial resolutions, we observed the expected convergence of the drift velocity towards a limiting value $v_\epsilon$ (Figure~\ref{fig:pinning}, right). Moreover, the computed $v_\epsilon$ values were found to be proportional to $\epsilon$. We attribute this agreement with the expected behavior to the fact, that in the LDG model, the additional potential on $\boldsymbol\theta$ (together with the square gradient term $|\nabla\boldsymbol\theta|^2$) introduces a new length scale of the orientation field over which the defects are regularized. Below this length scale, refining the grid resolution further correspond to a finer discretization of the \emph{same system}.

\subsection{Large-scale simulations}
\label{sec:largescalesims}

Finally we present real-world examples, i.e., large scale grain growth simulations that show the existence and effects of the defects in the HMP model and illustrate how they are cured by the new models. We first display selected parts of simulations providing a direct visual comparison of the critical regions. Then we show how the statistics of the grains are affected by these defects.
 
To this purpose we have simulated the models on a $4096 \times 4096$ pixels domain with periodic boundary conditions. The simulations were done in two stages. First, we simulated the \emph{solidification} of an undercooled ($u\lambda=0.5$) liquid. To mimic the orientational disorder in the liquid state the orientation field was set to uncorrelated random values in each pixel of the domain~\cite{Granasy2002}. Solidification was initiated by placing about 2300 small solid seeds with random orientation in the simulation box. No further noise was added to the system. Once solidification completed, the resulting multi-grain structure was used as initial condition for the subsequent \emph{grain growth} simulations. In most cases we used the same model to simulate for both stages, but in some cases, for the sake of easier comparisons, we used different models to simulate solidification and grain growth.

\subsubsection{Multi-grain structures with the HMP model}

We show two simulations to illustrate the behavior of the defects during grain coarsening. They differ in the multi-grain structure used as initial conditions for the grain growth simulations with the HMP model. In the first case, this structure was obtained by simulating solidification by the HMP, while in the second case with the LDG model. The main difference between them was in the number of defects along grain boundaries. Solidification with the HMP model produced lots of defects, especially along grain boundaries with misorientation $\Delta\theta \simeq \pi$, while with the LDG model the grain boundaries were practically defect-free. We explain this as follows. Due to the initially random orientations in the liquid phase, when orientational ordering takes place between two grains just about to impinge, both types of solutions can form along the same grain boundary. In the HMP model, due to the topological reasons and lattice pinning, the higher energy solutions cannot relax to the lower energy ones, while in the LDG model these defects relax easily.

\begin{figure}
\includegraphics[width=\linewidth]{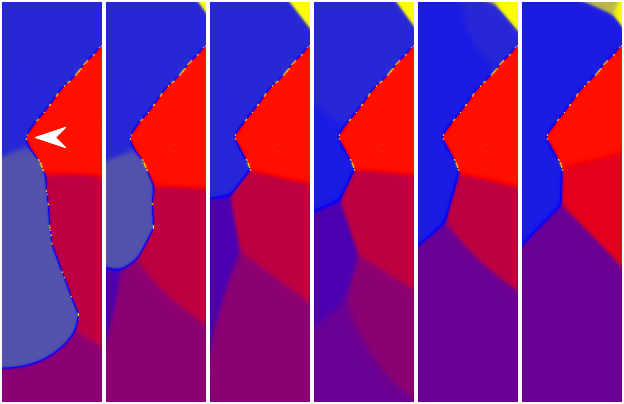}
\caption{Snapshots of the orientation field corresponding to 0, 5, 10, 20, 30 and 40 equal units of time of a grain coarsening simulation with the HMP model. The multi-grain configuration the simulation was started with was obtained by simulating solidification with the same HMP model. All snapshots show the same $100 \times 400$ pixel region of the full $4096 \times 4096$ pixel simulation domain. The color coding is the same as in Figure~\ref{fig:defect1c}. The defects, which are the points where the purple and yellow segments of the grain boundary (the two different solutions) meet, are mobile on the smaller angle grain boundary on the lower half of the snapshots. However, the defects on the larger angle grain boundary on the upper half are pinned and do not allow the grain boundary to move, resulting in an unphysical kink (shown by the white arrowhead) on the red-blue grain boundary.}
\label{fig:pinning1c1}
\end{figure}

Snapshots from the first simulation are shown in Figure~\ref{fig:pinning1c1}. We see a large number of defects on each picture, some of them are pinned by the simulation grid. As an extreme illustration of this pinning, a stable, sharp, unphysical kink can be seen on the snapshots.

\begin{figure}
\includegraphics[width=\linewidth]{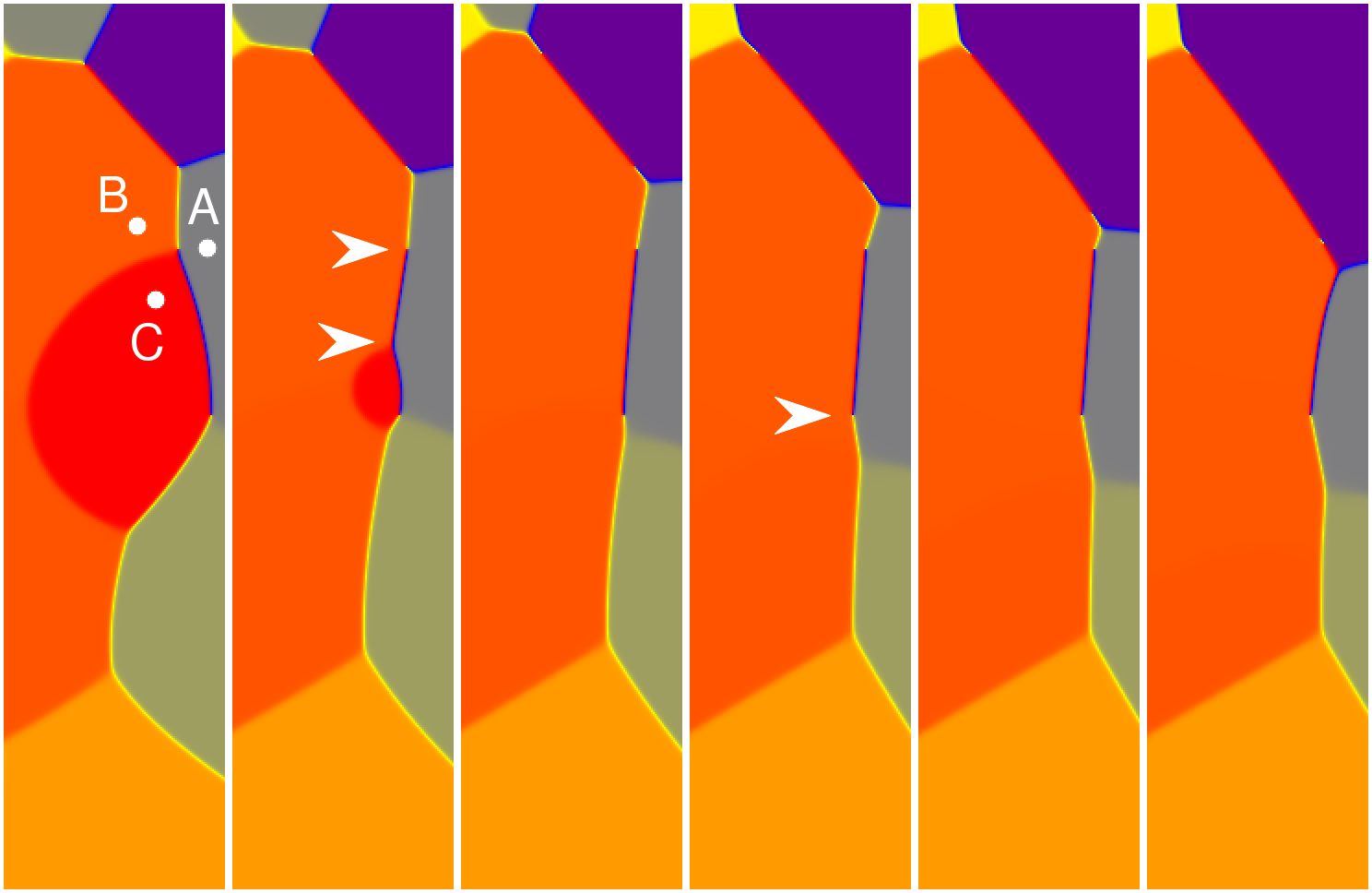}
\caption{Snapshots of the orientation field corresponding to 0, 1, 2, 4, 6 and 8 equal units of time of a grain coarsening simulation with the HMP model. To start from a defect-free configuration, the initial multi-grain structure was obtained by using the LDG model when simulating solidification. All snapshots show the same $120 \times 480$ pixel region of the full $4096 \times 4096$ pixel simulation domain. The color coding is the same as in Figure~\ref{fig:defect1c}. The signed misorientations between the grains marked by the white circles A, B and C are 0.37 (B--A) 0.13 (C--B) and slightly below 0.5 (A--C). Thus, the winding number corresponding to the enclosed trijunction on the leftmost snapshot is 1. The second snapshot shows that as the red grain shrinks, the trijunction point (defined as the meeting point of the orange, red and gray grains) is decoupled from the defect (defined as the point with winding number 1) and the defect stays visible long after the original red grain disappeared. The white arrowhead on the fourth snapshot shows a defect pinned by the lattice, being responsible for the kink in the grain boundary.}
\label{fig:pinning1c2}
\end{figure}

The snapshots from the second simulation (Figure~\ref{fig:pinning1c2}) are much smoother. Initially, defects exist only at trijunction points. However, as a result of pinning in the HMP model, the defects can be decoupled from the trijunction points. This illustrates that defects may not only be annihilated, but also created along grain boundaries in the HMP model during the course of grain growth simulations. Therefore their existence cannot be considered as being only a transient issue.

\subsubsection{Multi-grain structures with the US and LDG  models}

\begin{figure}
(a)~\includegraphics[width=0.9\linewidth]{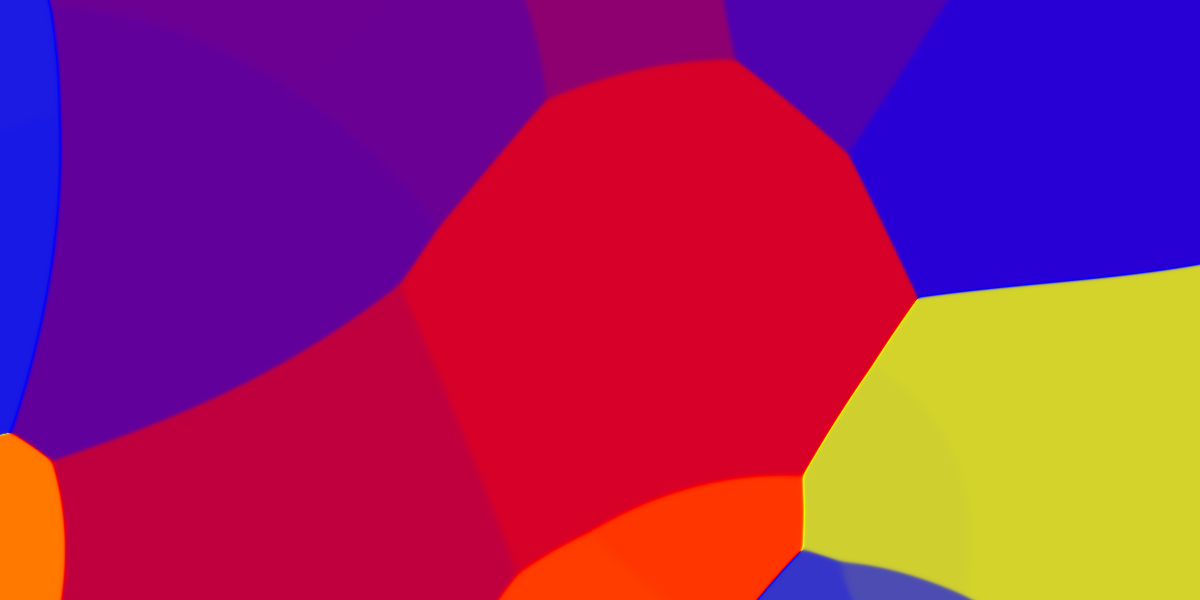}
(b)~\includegraphics[width=0.9\linewidth]{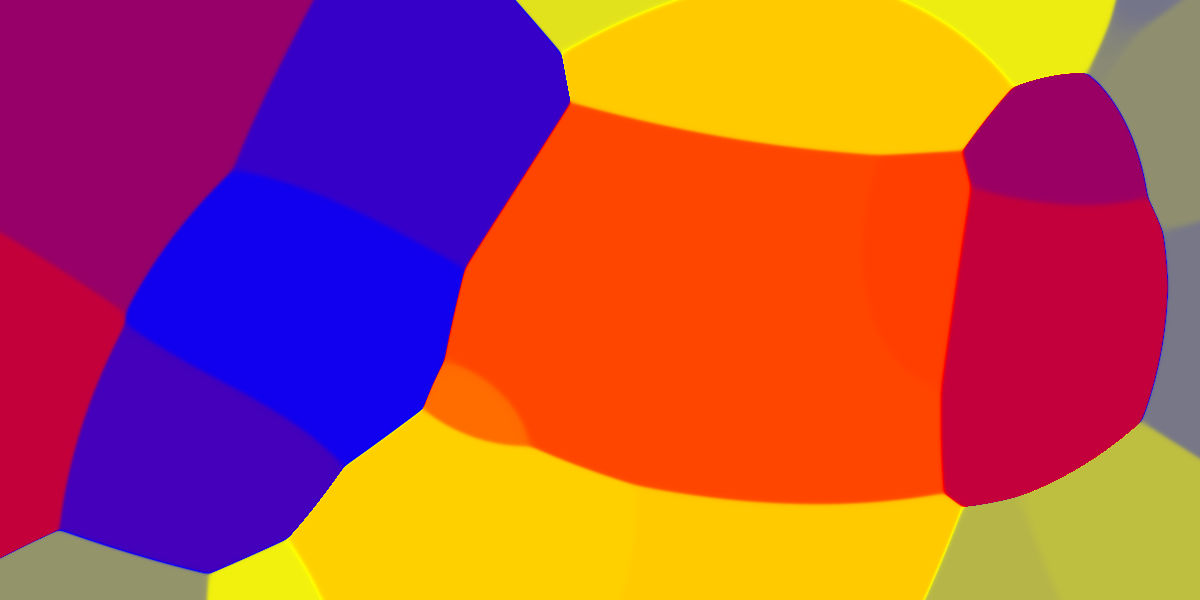}
\caption{Orientation field maps from simulations by the US (a) and LDG (b) models. The snapshots show randomly chosen $1200 \times 600$ pixel parts of the total $4096 \times 4096$ pixel simulation domain from a later stage of the simulation, when grains are relatively large.  The color coding is the same as in Figure~\ref{fig:defect1c}. No defects can be seen on the grain boundaries.}
\label{fig:newmodels}
\end{figure}

Figure~\ref{fig:newmodels} shows the scalar orientation field $\theta$ obtained as the polar angle from the components $\theta_1$ and $\theta_2$ of the US and LDG models. All grain boundaries are smooth and defect free. No blocking/pinning of the grain boundaries can be observed. Please notice that the high angle grain boundaries look very sharp in the $\theta$ map, but this is just the result of the $\theta_1, \theta_2 \rightarrow \theta$ conversion, the original order parameters are still well resolved.

\subsubsection{Comparison of GBDCs and LGSDs}

In the previous subsection we have pinpointed small regions of the simulation domain where we could see that the defects, especially the pinning of these defects modify the grain boundary dynamics. A practically important question is that to what degree do they modify the coarsening on the scale of the whole simulation. To check this, we evaluated two kinds of distributions which are frequently used when comparing multi-grain structures. One is the grain boundary character distribution (GBCD), which is the distribution of the relative length of all interfaces with a given misorientation. The other is the limiting grain size distribution (LGSD), which is the long-time steady-state distribution of the normalized grain size. We have evaluated both distributions the same way as described in Ref.~\cite{Korbuly2017a,Korbuly2017b}. 

\begin{figure}
\centerline{\includegraphics[width=\linewidth]{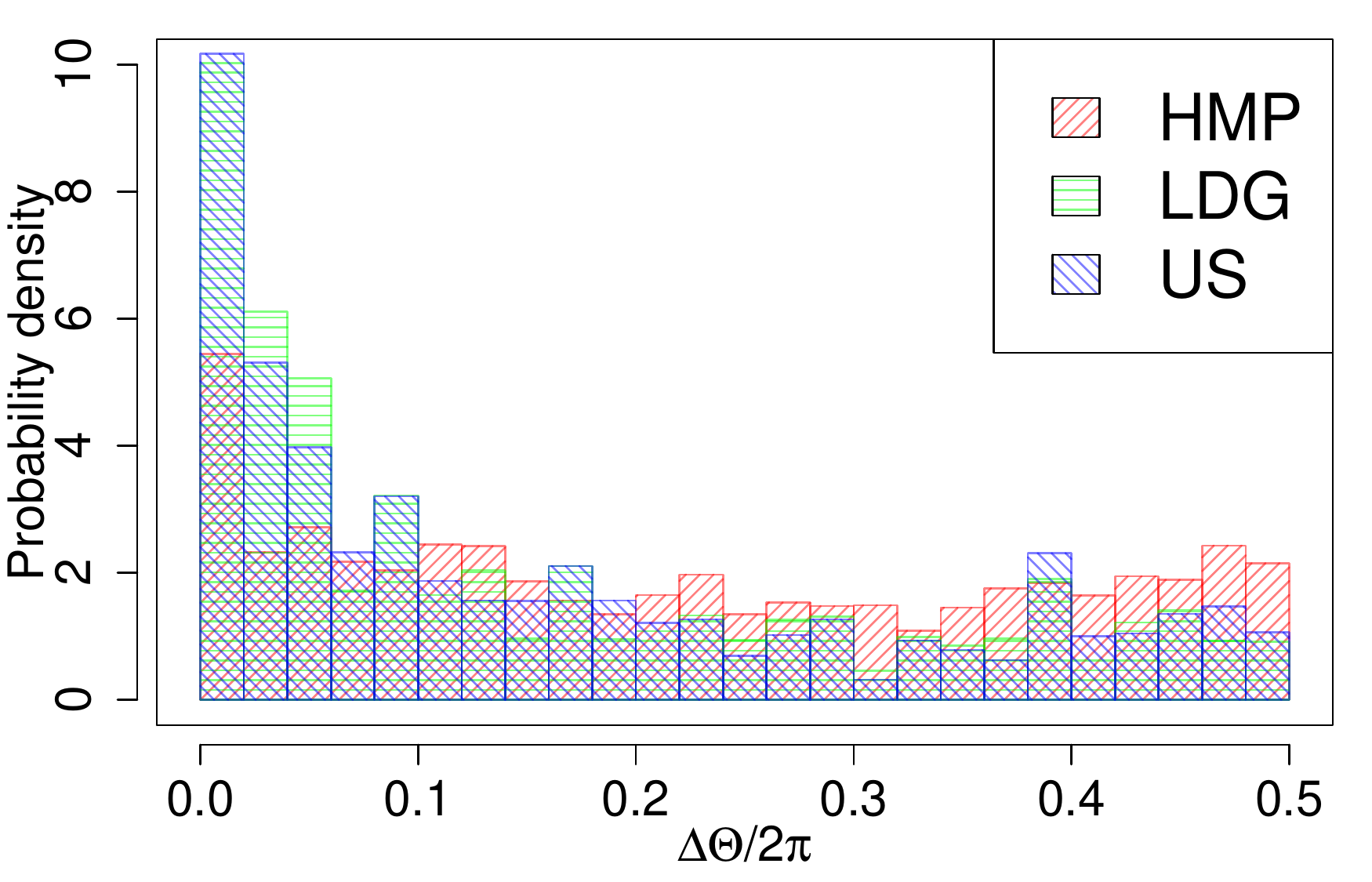}}
\caption{Grain boundary character distributions obtained by the different models. The length of grain boundaries with large misorientation is significantly larger in the HMP model than in the US and LDG models. We attribute this to the pinning of the defects which is most effective on grain boundaries with misorientation $\Delta\theta=\pi$.}
\label{fig:gbdc}
\end{figure}

Figure~\ref{fig:gbdc} shows that the GBCDs of the new models are very similar, but differ clearly from the GBDC of the HMP model. The relative length of the high angle grain boundaries in HMP model is higher than in the other models. This can be attributed to the longer life of the high angle grain boundaries due to their pinning by the topological defects. 

\begin{figure}
\centerline{\includegraphics[width=\linewidth]{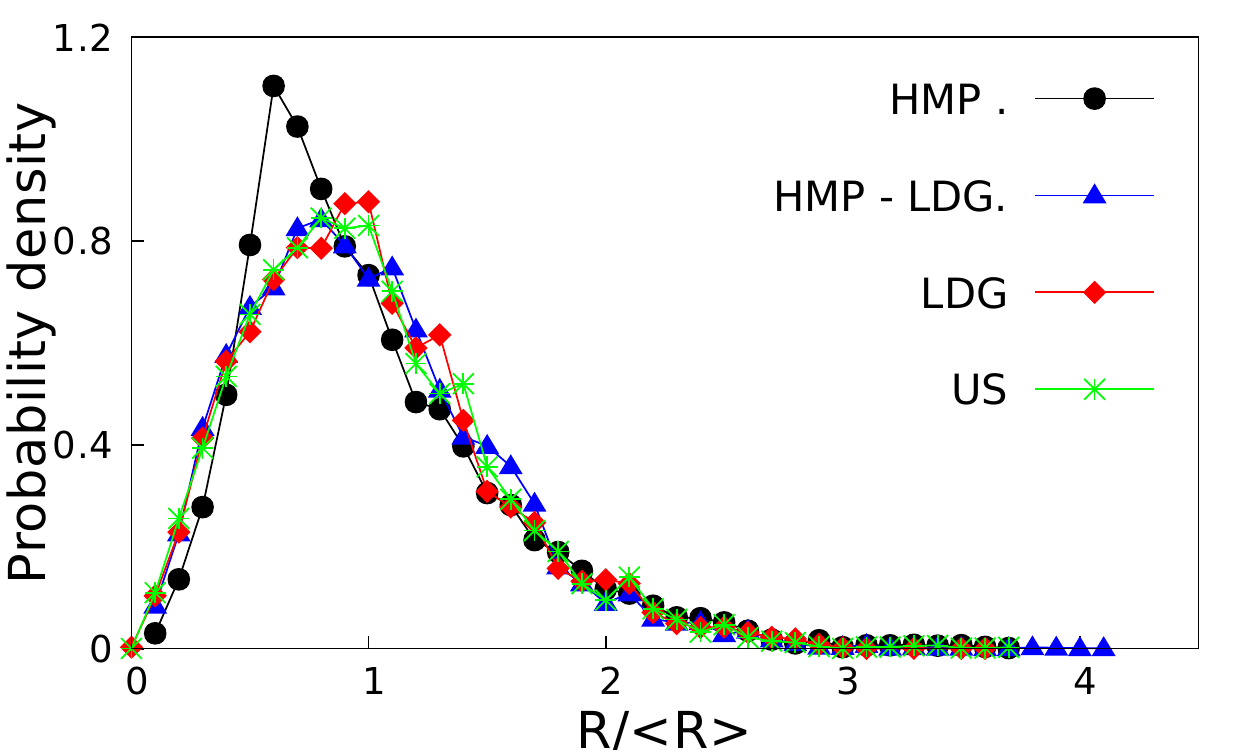}}
\caption{Limiting grain size distributions obtained by the different models. The new line labelled as HMP-LDG represent a variant of the HMP simulation, where the initial configuration of the HMP model was taken from the LDG simulation.}
\label{fig:lgsd}
\end{figure}

The LGSDs obtained by the different models are shown in Figure~\ref{fig:lgsd}. Just as in ~\ref{fig:gbdc}, the HMP result is clearly different from the US and LGD results. This suggests that the presence and pinning of defects can have a significant effect on the LGSD as well. For further insight, we added the LGSD from a new variant of the HMP model (labeled as HMP-LDG) in the figure. In this simulation we used the same defect-free initial multi-grain structure as used with the LDG model. The good agreement of this distribution with the US and LDG results and its clear difference to the HMP result suggests, that the large difference in the LGSD is due to the large number of defects formed during solidification in the HMP simulations. The effect of defects forming during grain growth seems to be negligible.

\section{Summary}
\label{sec:summary}

Based on general topological considerations, we made a detailed investigation of continuous solutions that can be obtained for multi-grain structures by using a scalar $\theta(\mathbf{r})$ orientation field. We identified two related phenomena that naturally occur in large-scale simulations of grain coarsening and that are difficult to interpret within a classical sharp interface or atomistic picture of grain boundaries: the existence of two different grain boundary solutions and topological defects on grain boundaries. We have shown that these singular defects on grain boundaries may be pinned by the grid used in numerical simulations, blocking the movement of the grain boundary. This blocking has visible effect on the results of large-scale simulations, as shown by the respective grain boundary character distributions and limiting grain size distributions. We have to note, however, that adding noise to the scalar orientation field can also un-pin these defects, resulting in LGSDs that are in good agreement with the ones produced by the new models. We found it worthwhile, however, to construct new models that are free of these defects by their nature.

Having realized that these problems originate in the topological properties of the scalar orientation field, we proposed two new models with new order parameters representing the 2D orientation. We have shown, first by focusing on the problematic details, then by carrying out large scale simulations that both of these new models are nonsingular and capable of circumventing the two problems identified. Though the two models differ considerably in their mathematical formulation, they are very similar not only in terms of their results, but also in their difficulty of implementation and numerical performance. The LDG model offers some flexibility by the possibility of adjusting the potential strength $\nu$, the US model may have the advantage of being equivalent to the original HMP model except for the neighborhood of trijunctions and defects, leaving e.g.\ the grain boundary energies unaffected.

It is important to note, however, that the new models are non-singular only in their original order parameter. When we interpret their results as the ``true'' scalar crystallographic orientation, singularities may re-appear. There are two important points, though. First, singularities appear only in places where necessary, i.e., only at trijunctions, and not along grain boundaries. Second, in contrast to the original model, the time evolution of the system is based on the new, non-singular order parameter. This means that we do not have to deal with singular fields during the numerical solution. Multi-phase-field models are very similar in this sense. There, the time evolution of the system is described by a set of non-singular phase-field variables. However, if we derive an orientation field as a weighted average of the individual orientations assigned to the phase-fields, we obtain a scalar field with singular points.

Finally, let us make a comment regarding the possible use of a true 3D orientation field (as opposed to the 3-component US model) in the 2D simulations. Having seen that the additional degree of freedom provided by the US and LDG models result in the disappearance of defects, it is tempting to think that using a 3D orientation field would have the same effect. As even thin layers of polycrystals, which can be considered as 2D samples, consist of real materials with true 3D crystal structure, this would be a nice physical escape from the problems related to the 2D orientation field. But unfortunately, this is not the case. The order parameter space of 3D orientations is also not simply connected~\cite{Costa2012}. Indeed, it is usually considered as either the full 3D sphere or the surface of the 4D sphere, both with antipodal points equated. This means that even using a true 3D orientation field, such as in Ref.~\cite{Pusztai2005,Kobayashi2005}, the two different grain boundary solutions and the associated topological defects would still exist. We have to note, however, that these defects could also be removed via similar treatments that we offered for the 2D orientation field. The quaternion representation of the rotation group uses 4D vectors of unit modulus. The Landau--de Gennes approach of relaxing this hard constraint and replacing it by an additional potential seems to be a promising alley for further research. But this, together with the increase of the spatial dimensions to 3, which result in a much greater variety of defect structures, would increase the complexity of the subject significantly. Therefore we leave the exploration of the 3D case for a possible future work.

\begin{acknowledgements}
This work was supported by the Hungarian-French Bilateral Scientific and Technological Innovation Fund under Grant no.~TÉT\_12\_FR-2-2014-0034; and the National Agency for Research, Development, and Innovation (NKFIH), Hungary under Contract no.~OTKA-K-115959.
\end{acknowledgements}

\appendix
\section*{Appendix}

Let us assume that orientation is represented by an $N$-component unit vector, i.e.~$\boldsymbol{\theta}=(\theta_1,\ldots,\theta_N)$ with the constraint
\begin{equation}
\label{eq:constraint1}
\boldsymbol\theta^2 = \sum_{i=1}^N \theta_i^2 = 1.
\end{equation}
This means that the order parameter space is the $N\!-\!1$ dimensional hypersurface of the $N$ dimensional unit sphere, or shortly the $N\!-\!1$-sphere. In this Appendix we derive the equation of motion for $\boldsymbol\theta$ that maintains this constraint by ensuring that the length of $\boldsymbol\theta$ does not change by time.
The formal expression of this requirement is 
\begin{equation}
\label{eq:constraint2}
\boldsymbol\theta \dot{\boldsymbol\theta} = \sum_{i=1}^N \theta_i \dot\theta_i = 0,
\end{equation}
which is the time derivative of Eq.~\ref{eq:constraint1}.

Let us consider a general free energy functional
\begin{equation}
F[\phi,\{\theta_i\}] = \int f(\phi,\nabla\phi,\{\theta_i\},\{\nabla\theta_i\})\,dV.
\end{equation}
If the $\theta_i$-s were unconstrained then the variational approach would result in the standard Allen-Cahn equations of motion
\begin{align}
\label{eq:standardeom}
\dot\theta_i = -M_\theta \frac{\delta F}{\delta \theta_i} = -M_\theta \left( \frac{\partial f}{\partial\theta_i} - \nabla\frac{\partial f}{\partial\nabla\theta_i} \right),
\end{align}
assuming a common mobility $M_\theta$ for all components. By using these \emph{non-constrained} equations of motion, $\boldsymbol\theta$ would not remain a unit vector, it would be driven off the $N-1$-sphere. To derive the \emph{constrained} equations of motion that obey Eq.~\ref{eq:constraint2} and therefore Eq.~\ref{eq:constraint1} we use the standard Lagrange multiplier method.

First, we construct a modified free energy density and free energy functional by adding a new term containing the constraint, 
\begin{align}
f^* = f + \Lambda \left( 1-\sum_{i=1}^N \theta_i^2 \right)
\end{align}
and
\begin{align}
F^* = \int f^* \, dV,
\end{align}
where $\Lambda$ is the unknown Lagrange multiplier. Then, starting from these modified expressions, the standard derivation (Eq.~\ref{eq:standardeom}) result in the conserved equations of motion
\begin{align}
\label{eq:constrainedeom1}
\dot\theta_i^* &= -M_\theta \frac{\delta F^*}{\delta \theta_i} = -M_\theta \left( \frac{\delta F}{\delta \theta_i} - 2 \Lambda \theta_i \right),
\end{align}
which contain the unknown Lagrange multiplier. We can obtain the extra equation required to determine $\Lambda$ by multiplying Eq.~\ref{eq:constrainedeom1} by $\theta_i$ and summing over all $i$-s,
\begin{align}
\label{eq:determinelambda1}
\sum_{i=1}^N \theta_i \dot\theta_i^* = -M_\theta \left( \sum_{i=1}^N \theta_i \frac{\delta F}{\delta \theta_i} - 2 \Lambda \sum_{i=1}^N \theta_i^2 \right),
\end{align}
which simplifies to 
\begin{align}
\label{eq:determinelambda2}
2 \Lambda = \sum_{i=1}^N \theta_i \frac{\delta F}{\delta \theta_i}
\end{align}
because of Eq.~\ref{eq:constraint2} and Eq.~\ref{eq:constraint1}. Plugging this back to Eq.~\ref{eq:constrainedeom1}, we arrive to the final form of the constrained equations of motion,
\begin{align}
\begin{split}
\label{eq:constrainedeom2}
\dot\theta_i^* = -M_\theta \left( \frac{\delta F}{\delta \theta_i} - \theta_i \sum_{k=1}^N \theta_k \frac{\delta F}{\delta \theta_k} \right) = \\
= \dot\theta_i - \theta_i \sum_{k=1}^N \theta_k \dot{\theta_k},
\end{split}
\end{align}
or, using the $N$-dimensional vector notation,
\begin{align}
\label{eq:constrainedeom3}
\dot{\boldsymbol\theta}^* = \dot{\boldsymbol\theta} - \dot{\boldsymbol\theta}({\boldsymbol\theta}\dot{\boldsymbol\theta}),
\end{align}
where the $\dot\theta_i$-s defining $\dot{\boldsymbol\theta}$ are given by Eq.~\ref{eq:standardeom}.

This final form has a simple geometric interpretation. The $\dot{\boldsymbol\theta}^*$ constrained time derivative is obtained from the $\dot{\boldsymbol\theta}$ non-constrained time derivative with a projection to the $N\!-\!1$-dimensional plane which is tangential to the $N$-dimensional unit sphere at $\boldsymbol\theta$. Thus, $\dot{\boldsymbol\theta}^*$ is perpendicular to $\boldsymbol\theta$, satisfying the constraint Eq.~\ref{eq:constraint2}. Please note, however, that Eq.~\ref{eq:constraint2} guarantees Eq.~\ref{eq:constraint1} only for infinitesimal changes of $\boldsymbol\theta$. Any finite increment $\Delta\boldsymbol\theta = \dot{\boldsymbol\theta}^* \Delta t$ calculated in a numerical simulation for a time step $\Delta t$ violates Eq.~\ref{eq:constraint1} in the same way as any finite tangential movement causes a drift off a circle. We correct for this at the end of each time step by a radial projection which forces the incremented value of $\boldsymbol\theta$ back to the $N$-dimensional unit sphere.

\end{document}